\documentclass[aps,prb,twocolumn,superscriptaddress]{revtex4-1}

\usepackage{amsmath,amssymb} % for math symbols
\usepackage{physics} %mainly for braket notation
\usepackage{graphicx}% Include figure files
\usepackage{color}
\usepackage{dcolumn}% Align table columns on decimal point
\usepackage{bm}% bold math
\usepackage{siunitx}
\usepackage[mathlines]{lineno}% Enable numbering of text and display math
\begin{document}

\title{Investigating the role of anion polarizability in Fe-based superconductors via light-matter interaction}
\author{Xiaoxiao Xiong}
\affiliation{\!Department \!of \!Physics and Astronomy, \!University
	of\!  British Columbia, \!Vancouver, British \!Columbia,\! Canada,\!
	V6T \!1Z1}
\affiliation{\!Stewart Blusson Quantum Matter \!Institute, \!University
	of British Columbia, \!Vancouver, British \!Columbia, \!Canada,
	\!V6T \!1Z4}
 \affiliation{Department of Physics and Institute for Quantum Information and Matter, California Institute of Technology, Pasadena, California, USA 91125}
\author{Fabio Boschini}
\affiliation{\!Stewart Blusson Quantum Matter \!Institute, \!University
	of British Columbia, \!Vancouver, British \!Columbia, \!Canada,
	\!V6T \!1Z4}
\affiliation{Centre \'{E}nergie Mat\'{e}riaux T\'{e}l\'{e}communications, Institut National de la Recherche Scientifique, Varennes, Qu\'{e}bec J3X 1S2, Canada}

\author{Mona Berciu}
\affiliation{\!Department \!of \!Physics and Astronomy, \!University
	of\!  British Columbia, \!Vancouver, British \!Columbia,\! Canada,\!
	V6T \!1Z1}
\affiliation{\!Stewart Blusson Quantum Matter \!Institute, \!University
	of British Columbia, \!Vancouver, British \!Columbia, \!Canada,
	\!V6T \!1Z4}
\date{\today}

\begin{abstract}
The polarizability of nearby ions may have a significant impact on electron interactions in solids, but only limited experimental data are available to support this picture. In this work, using a highly simplified description of the prototypical FeAs superconducting layer, we show how external optical excitation of the As 4p-5s splitting can lead to a significant modulation of the polarization-mediated effective interactions between carriers. Our results suggest that even perturbative external fields, approximately two orders of magnitude smaller than the internal field generated by charge carriers, might enable the exploration of the role of the anion's polarizability in determining the correlated physics, although more detailed modeling is needed to decide optimal ways to achieve this.
\end{abstract}
\maketitle

\section{Introduction}

There are many proposed theories of superconductivity (SC), ranging from the well established  Bardeen-Cooper-Schrieffer theory for phonon-mediated low-$T_c$ superconductors,\cite{BCS} to more speculative ideas such as bipolaronic superconductivity, \cite{Alexandrov,John1} the Little model for organic superconductors, \cite{Little}  the Allender, Bray, and Bardeen  excitonic models for superconductivity, \cite{ABB} the theory of hole superconductivity of Hirsch and co-workers,\cite{Hirsch1,Hirsch2,Hirsch3} and many variations on the theme of a superconducting glue arising from strong electron-electron correlations, initiated by Anderson \cite{Anderson} in the context of cuprate superconductivity.\cite{cuprate}

The underlying commonality of all these theories is that their  'glue' helps to (over)screen the bare Coulomb repulsion between charge carriers, resulting in an effective attraction  that favors the formation of Cooper pairs. This suggests that multiple 'glues' could contribute to superconductivity in one material, raising the question of how to identify and differentiate their relative contributions. For phononic and correlation-based 'glues' (which are the basis of all theories mentioned above) this issue has been investigated by different means, \cite{SJ1, SJ1b, SJ2, SJ3} although more work is needed.

In this work, we focus on a less studied 'glue', which could arise from the polarizability of nearby ions and acts analogously to more established 'glues' (e.g. phonons) as a source of screening. For specificity, we focus on its role in Fe-based superconductors, where this mechanism was first proposed\cite{george1} in the context of SC  and was argued to contribute significantly to  (over)screening. \cite{george1, george2} We emphasize, however, that this mechanism contributes to screening in any material. Furthermore, we propose an experimental strategy employing light-matter interactions to assess the relevance of non-uniform atomic polarizability in determining the electronic properties of FeAs. This approach aligns with recent theoretical \cite{valmispild, tancogne-dejean, golez} and experimental \cite{baykusheva, beaulieu} studies that have reported light-induced modification of the atomic on-site Coulomb interaction.

\begin{figure}[t]
    \centering  \includegraphics[width=\columnwidth]{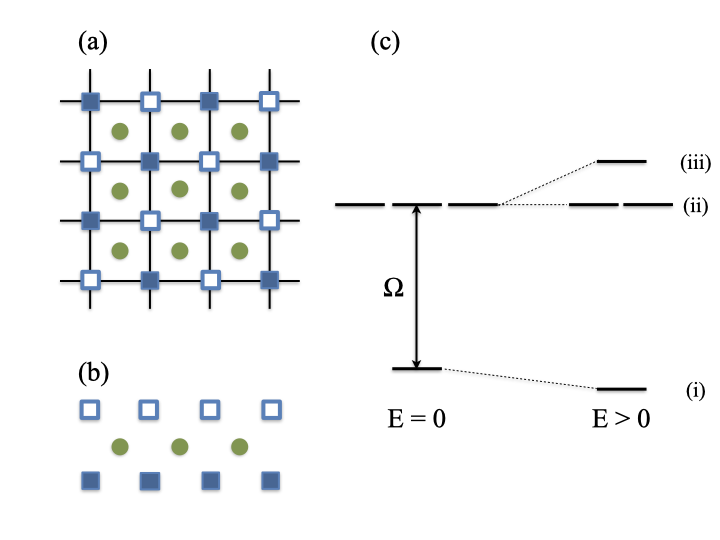}
    \caption{FeSe layer as seen from above (a) and sideways (b). Green circles show Fe locations, while empty/full blue squares show  As ions located above/below the Fe plane; (c) spectrum of an unpolarized (left) and a statically polarized (right) As ion. See text for more details. }
    \label{fig1}
\end{figure}

The main idea behind our work can be explained in terms of basic electromagnetism. Consider a FeAs layer, which has Fe on a square lattice and As occupying alternating corners of the cubes that have Fe at their centers, such that half of the As are above, and half are below the Fe layer, see Fig. \ref{fig1} a,b.  Note that we ignore distortions of the cubes; they are trivial to incorporate in the analysis but lead to small corrections to the estimates provided below.

The electronic levels with As character are far from the Fermi energy, hence the As anions are seemingly not relevant to the electronic properties. However, these large anions have significant polarizability $\alpha_p$ and develop dipole moments $\langle{\bf p}\rangle = \alpha_p {\bf E}$ when subject to the electric field ${\bf E}$ created by additional carriers located on Fe orbitals. For a single carrier, this implies a lowering of its on-site energy by $- \alpha_p \abs{\bf E}^2/2$ (at a semi-classical level) for each polarized As, where the field ${\abs{\bf E}}$ depends on the Fe-As distance. This energy lowering results in the formation of a quasiparticle dressed by this polarization cloud -- the so-called electronic polaron -- with a renormalized band-structure and an enhancement in the effective mass (by about a factor of 2 in FeAs\cite{george2}).

As typical in polaronic physics, the clouds affect not only the quasiparticle dispersion but also the effective interactions between quasiparticles: if two carriers are sufficiently close to one another, they interact with each other's clouds,  thereby changing the total polarization energy. Specifically, an As that experiences electric fields ${\bf E}_1$ (${\bf E}_2$) from carrier 1 (2) contributes $- \alpha_p \abs{{\bf E}_1+ {\bf E_2}}^2/2$ to the total energy. Of these three terms, $- \alpha_p \abs{\bf E_1}^2/2$ and $- \alpha_p \abs{\bf E_2}^2/2$ are associated with the individual electronic polarons, while $-\alpha_p {\bf E_1}\vdot {\bf E_2}$ is an effective interaction between the two polarons, with a sign and magnitude that depend on the relative distances between carriers and between carriers and the As ion involved. In particular, if both carriers are on the same Fe then ${\bf E}_1 ={\bf E_2}$ and the effective on-site interaction $-\alpha_p \abs{\bf E_1}^2$ is attractive, resulting in  a substantial screening of the on-site Hubbard repulsion by up to  5-10 eV (see below). For carriers further apart, the magnitude of this effective interaction is smaller but may still contribute significantly to screening and thus influence the appearance of SC. 

Here we investigate whether it is possible to quantify the relevance of this screening mechanism in Fe-based superconductors via tailored light excitation. At a microscopic level, an external field ${\bf E}$ can induce a dipole moment by exciting electrons from the full $4p$ shell primarily into the empty $5s$ shell of As$^{3-}$. A sufficiently intense external electric field with a frequency close to the $4p-5s$ resonance can partially saturate these transitions, thus interfering with the polarization induced by carriers, altering the strength of their effective interactions, and ultimately affecting the electronic properties of the system. The main goal of this article is to understand whether the modulations due to an external pump could be significant enough to make such investigations feasible.  As discussed below, we believe that our results are encouraging, motivating the study of much more detailed models in future work. 

The article is organized as follows: Section II introduces the model and its parameters, Section III presents our results in the absence of the pump, Section IV presents results in the presence of the pump, and Section V contains our discussion of the results.

\section{The model}

% Introduce the lattice structure

As a first step in answering the  question stated above, we strip down the model used in Ref. \onlinecite{george2} to keep only the terms most relevant to the As polarization (the approximations are listed in this section and further discussed in the last section), allowing us to carry out simulations relatively easily in order to get rough estimates for the magnitudes of various energy modulations.

First, we assume that there is a single valence orbital per Fe site and ignore its band dispersion so that charges placed on Fe ions remain fixed in space. As a result, these charges are described by:
%%%%%%%%%%%%%%%%%%%%%%%%%%%%%% EQUATION %%%%%%%%%%%%%%%%%%%%%%%%%%%%%%
\begin{equation}
\mathcal{H}_{Fe} = U_H \sum_{j \in Fe}\hat{n}_{j\uparrow} \hat{n}_{j\downarrow}. \label{eq: H_Fe}
\end{equation}
%%%%%%%%%%%%%%%%%%%%%%%%%%%%%%%%%%%%%%%%%%%%%%%%%%%%%%%%%%%%%%%%%%%%%%
where $\hat{n}_{j\sigma}$ counts the number of charges with spin $\sigma$ at site $j$ of the Fe square lattice. $U_H$ is the on-site Hubbard repulsion between these charges. \textcolor{black}{It includes screening from all other mechanisms {\em apart} from As polarizability (the latter effect is considered explicitly below), meaning that here $U_H$ is significantly larger than in models where the As have already been integrated out. }The other screening mechanisms are assumed to make longer-range interactions vanishingly small, although finite longer-range repulsion can be introduced in the model trivially.

Second,  we follow the model in Ref. \onlinecite{george2} and describe the As not in terms of electrons excited from $4p$ into $5s$, but in terms of holes excited from $5s$ into $4p$ orbitals. Thus, for each As ion we introduce the {\em hole} annihilation operators $s_\sigma, p_{\lambda,\sigma}$, where $\lambda=x,y,z$ indicates the direction of the $p$ orbitals. In hole language, the ground state of an isolated As$^{3-}$ ion has the $5s$ orbital completely occupied with holes (empty of electrons) while $4p$ is completely empty of holes (fully occupied by electrons), therefore:
%%%%%%%%%%%%%%%%%%%%%%%%%%%%%% EQUATION %%%%%%%%%%%%%%%%%%%%%%%%%%%%%%
\begin{equation}
 \mathcal{H}_{As} = \Omega \sum_{i \in As,\lambda, \sigma} p_{i,\lambda,\sigma}^\dagger p_{i,\lambda,\sigma}
    \label{eq: H_As}
\end{equation}
%%%%%%%%%%%%%%%%%%%%%%%%%%%%%%%%%%%%%%%%%%%%%%%%%%%%%%%%%%%%%%%%%%%%%%
where $\Omega>0$ is the energy spliting between the $4p$ and $5s$ shells (we set $\hbar=1$). The finite bandwidth of these bands  is also ignored as in Ref. \onlinecite{george2}. 

The polarization of the As ions due to charges located on the Fe ions is described by:
\begin{align}
    \mathcal{H}_{p} &= \sum_{j\in Fe} \hat{n}_j \sum_{i \in As,\sigma} g_{ij} \big( s_{i,\sigma}^\dagger \bar{p}_{ij,\sigma} + h.c. \big).
    \label{eq: interaction}
\end{align}
Here, $\hat{n}_j = \hat{n}_{j \uparrow} + \hat{n}_{j \downarrow}$ counts the number of carriers on the Fe site $j$. The coupling $g_{ij} =  \sqrt{\alpha_p \Omega}E_j(i)/2$ is controlled by the magnitude of the electric field ${\bf E}_j(i)$ created by one carrier located on Fe$_j$ at As$_i$,  at a relative distance $r_{ij}$ in the direction of the unit vector ${\bf \hat{r}}_{ij}$. In the absence of screening  ${\bf E}_j(i) = {\bf \hat{r}}_{ij} e/r_{ij}^2$; screening (from other sources) decreases the magnitude and range of ${\bf E}_j(i)$. Finally, the annihilation operator $\bar{p}_{i,j,\sigma} = \sum_{\lambda}^{}({\bf \hat{r}}_{ij} \cdot {\bf \hat{e}}_\lambda) p_{i,\lambda,\sigma}$ removes a hole from the $4p$ orbital parallel to  ${\bf \hat{r}}_{ij}$. The occupation of the two  $4p$ orbitals perpendicular to ${\bf \hat{r}}_{ij}$ is not affected by the electric field.

To model the interaction with the external electric field ${\bf E}_a$ (e.g. produced by a continuous wave laser), we use a similar term:
\begin{align}
    \mathcal{H}_{ext}(t) &= g_a \cos(\omega_a t) \sum_{\sigma, i \in As} (s_{i,\sigma}^\dagger \bar{p}_{ia, \sigma} + h.c.).
    \label{eq: H_laser}
\end{align}
We assume a monochromatic laser instead of an ultrashort pulse; the implications are discussed in the last section. The laser electric field ${\bf E}_a(t)= {\bf \hat{\epsilon}}_a E_a\cos(\omega_a t)$ is polarized parallel to the unit vector ${\bf \hat{\epsilon}}_a$ and oscillating with a  frequency $\omega_a$. The coupling constant to the laser electric field is  $g_a =  \sqrt{\alpha_p \Omega}E_a/2$ and the $4p$ orbital into which the laser can excite holes is  $\bar{p}_{ia,\sigma} = \sum_{\lambda}^{}({\bf \hat{\epsilon}}_a \cdot {\bf \hat{e}}_\lambda) p_{i,\lambda,\sigma}$. 

Therefore, the total Hamiltonian we consider is:
\begin{align}
     \mathcal{H}(t) = \mathcal{H}_{Fe} + \mathcal{H}_{As} + \mathcal{H}_{p} + \mathcal{H}_{ext}(t) \label{eq: H_tot}
\end{align}

% Choice of parameters
    Even after all these simplifications, the model is characterized by seven parameters: $\Omega, \alpha_p, g_1, g_2,   U_H, \omega_a, g_a$. Following Ref. \onlinecite{george2}, the $5s-4p$ energy difference is taken to be $\Omega = {6}$eV
    and the As polarizability is estimated to be $\alpha_p = 10-{12}{\AA}^3$. The couplings $g_1$ and $g_2$  are for the nearest-neighbor (nn) and second nearest-neighbor (nnn) As ions, respectively, located at a distance $r_1=a \sqrt{3}/2$ and $r_2 = a \sqrt{11}/2$.  In the absence of any screening, $g_2/g_1 = E_2/E_1 = 3/11$. Screening will further reduce this ratio, thus supporting our approximation of ignoring coupling between As and Fe that are at a distance $r > r_2$. In the absence of any screening and for $\Omega, \alpha_p$ values cited above, $g_1 \sim {2.5}$eV. The value of $U_H$ is irrelevant at this level of modeling, as discussed below. Finally, the frequency of the laser will be tuned across the resonance $\omega_a =  \Omega $, while the field strength is typically chosen as ${1-10}${MV/cm} leading to a typical $g_a \approx {0.05}$eV.

%%%%
\section{Results in the static case}

To set the stage, we first briefly review the results in the static case (no external perturbation, $g_a=0$).

\subsection{Spectrum of a polarized As}

Given any distribution of Fe charges, we calculate the resulting electric field ${\bf E}$ at any given As site, and its corresponding coupling $g = \sqrt{\alpha_p \omega} E/2$. The Hamiltonian for this As is then trivial to diagonalize and, as sketched in Fig. \ref{fig1}(c), for each spin projection it consists of three energy levels: (i) the ground-state with energy
$E_0(g) = {1\over 2}[\Omega - \sqrt{\Omega^2+4g^2}] <0$;
(ii) a doubly-degenerate state  at energy $\Omega$; (iii) the highest eigenstate with energy $E_1(g)= {1\over 2}[\Omega + \sqrt{\Omega^2+4g^2} ]> \Omega$. States (i) and (iii) are linear combinations of the $s$ and the $p_{\parallel}$ orbital oriented parallel to ${\bf E}$, while states (ii) are the two $p_{\perp}$ orbitals oriented perpendicularly to ${\bf E}$. (See Appendix A.)

Given its spin degeneracy, the contribution of this As ion to the ground-state energy is:
\begin{align}
    E_{As}(g) = 2 E_0(g) = \Omega - \sqrt{\Omega^2 + 4 g^2}.
    \label{eq: E_cloud}
\end{align}

\subsection{Energy of an electronic polaron}

We now assume that there is a single carrier \textcolor{black}{doped into the otherwise charge-neutral system. This carrier is located} on a Fe ion, and $g_1$ ($g_2$) is  the coupling to each of its 4 nn (8 nnn) As. The couplings are identical for all As at the same distance from the Fe because the electric fields have the same magnitudes, although they are pointing in different directions. Because the magnitude of the electric field decreases like $1/r^2$ (or faster, if there is screening) one can safely set to zero the coupling for As placed further than a cutoff distance. In Ref. \onlinecite{george2}, the cutoff was chosen such that only $g_1\ne 0$, here we instead set $g_1, g_2 \neq 0$ and show that the contribution from the 2nd coordination ring of As can be considerable if screening is completely ignored. Since screening is always present to some extent, we anticipate that the cutoff falls between the two extreme cases analyzed previously.

We ignore the small dipole-dipole interactions, hence the total energy of the polaron is the sum of energy contributions from each polarized As ion:
\begin{align}
    \begin{split}
        E_{P,GS}(g1,g2) &= 4 E_{As}(g_1) + 8 E_{As}(g_2). \label{eq: E_GS}
    \end{split}
\end{align}

\begin{figure}[t]
    \centering
    \includegraphics[width=0.9\columnwidth]{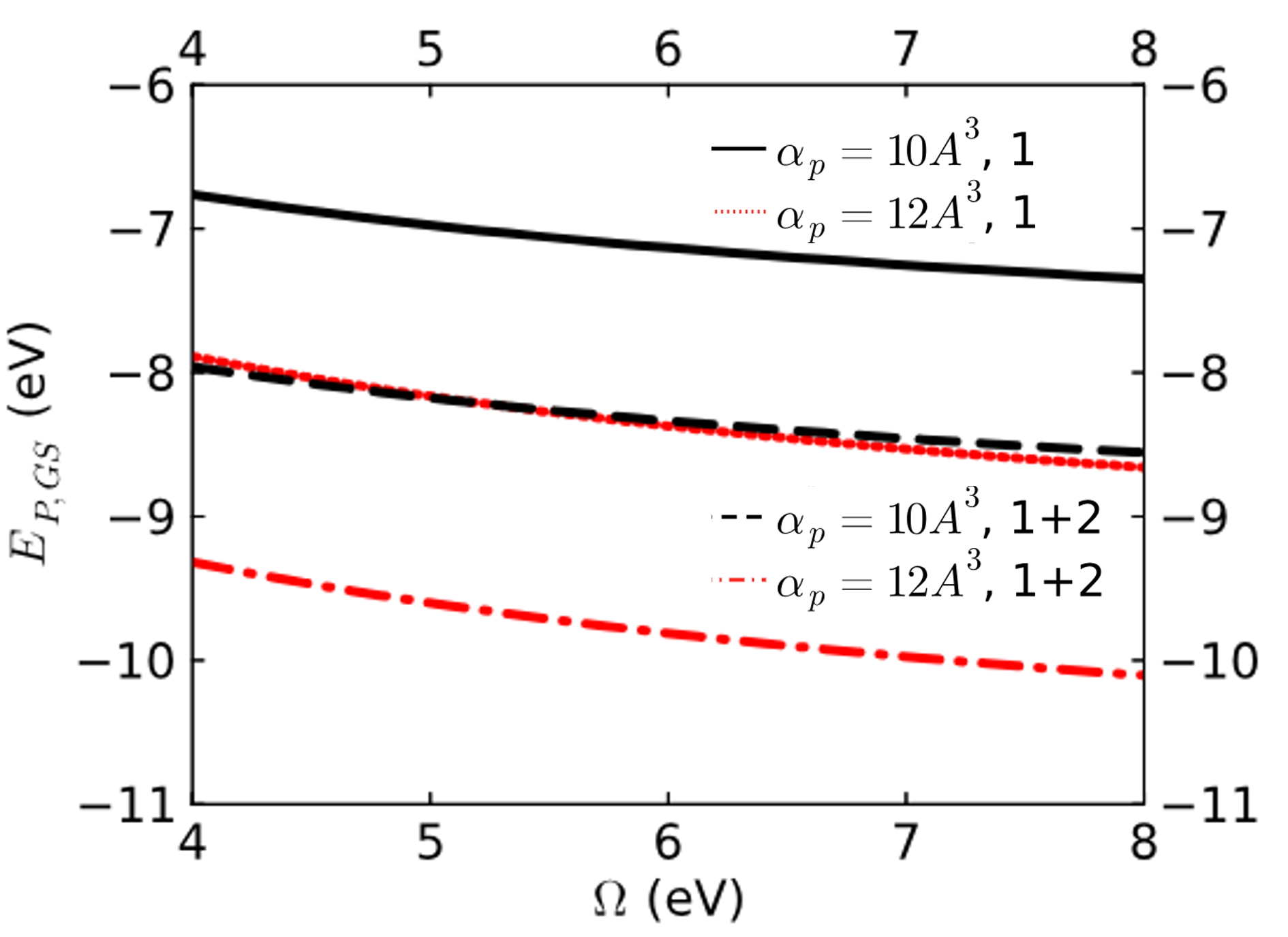}
    \caption{Total energy of a single polaron as a function of s-p transition energy $\Omega$, plotted for different polarizabilities $\alpha_p$. The label '1' refers to contributions from only the 4 nn As polarization clouds ($g_2 = 0$), while '1+2' also includes the 8 nnn As clouds. This plot assumes no screening.}
    \label{fig: E_pgs}
\end{figure}

We plot this energy as a function of $\Omega$ for two values $\alpha_p=10, 12 \AA^3$ in Fig. \ref{fig: E_pgs}. We see that depending on the parameters and the chosen cutoff for the cloud size, the creation of the polarization clouds can lower the energy of the carrier by anywhere from 7-10 {eV}. 

\subsection{Effective interaction between two polarons}
% Graph of polaron energies, effective attractions

\begin{figure*}[!t]
    \centering
\includegraphics[width=\linewidth]{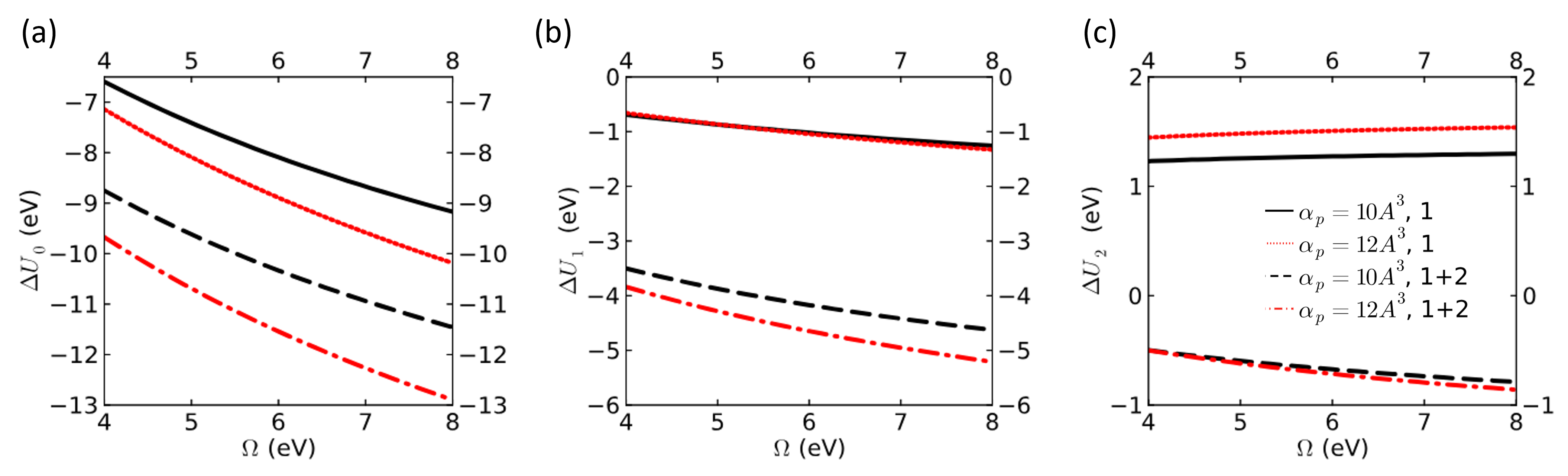}
    \caption{(a) Effective on-site attraction $\Delta U_0$; (b) Effective attraction between nearest-neighbour polarons $\Delta U_1$; (c) Effective attraction between second-nearest-neighbour  polarons $\Delta U_2$. Labels '1' and '1+2' means that contributions from only nn As and nn+nnn As, respectively, were included. These plots assume no screening. }
    \label{fig: polaron_energies}    
\end{figure*}

% What is effective attraction
Next, we consider the case with two charge carriers present in the Fe sublattice \textcolor{black}{of the otherwise charge-neutral system}. $E_n$ is the contribution from their polarization clouds when the two carriers are n$^\text{th}$ nearest neighbours. In the limit $n\rightarrow \infty$, each carrier  polarizes independently the As ions surrounding it, hence:
\begin{align}
    E_\infty(g_1,g_2) =  2E_{P,GS}(g_1,g_2).
\end{align}

When the carriers are close to each other, some As ions experience the combined electric field of both charges. This modifies their respective coupling $g$ accordingly, and thus their contribution to the total energy.

%The directions of $\bar{p}$ will also follow the direction of the induced dipole moment.

Specifically, if the two carriers are on the same Fe ($n=0)$, this corresponds to doubling the field experienced by each As polarization cloud, see Eq. (\ref{eq: interaction}). This is equivalent with scaling  $g_1 \rightarrow 2g_1, g_2 \rightarrow 2g_2$, and therefore
\begin{align}
    E_0 = E_{P,GS}(2g_1,2g_2).
\end{align}
The difference $\Delta U_0 =  E_0 - E_\infty<0$ defines an effective on-site attraction mediated by the polarization clouds. We plot $\Delta U_0$  for various parameters in Fig. \ref{fig: polaron_energies}(a), showing that it leads to a very substantial screening of the total on-site Hubbard energy  from $U_H$ to $U_0=U_H + \Delta U_0$. Whether $U_0$ is attractive or still repulsive \textcolor{black}{in our model} depends on the various parameters, \textcolor{black}{but in the actual materials it is believed to be moderately repulsive.  }

The effective polarization-mediated interaction is calculated similarly for all other configurations, with $\Delta U_n = E_n-E_{\infty}$. The range of this effective interaction depends sensitively on the distance cutoff for the included couplings. If this cutoff is set to only include nn As, then $\Delta U_n = 0$ for all $n\ge 3$.  If the cutoff includes also nnn As, then $\Delta U_n =0$ only for $n\ge 8$. 

Representative results for $\Delta U_0$, $\Delta U_1$ and $\Delta U_2$  are shown in Fig. \ref{fig: polaron_energies} for various parameters and cutoffs. Including the second coordination ring makes all these effective interactions significantly more attractive, in particular $\Delta U_2$ switches from repulsive into attractive. 

We note that we have also done these calculations including contributions from the third coordination ring of As ions. The variations in the effective attractions are less than $ {0.2}$eV for our typical parameters even when the electric fields are assumed to be completely unscreened. Since screening would reduce these contributions further, we ignore them throughout.

% Figure with contour plots
\begin{figure}[b]
        \centering
        \includegraphics[width=0.85\linewidth]{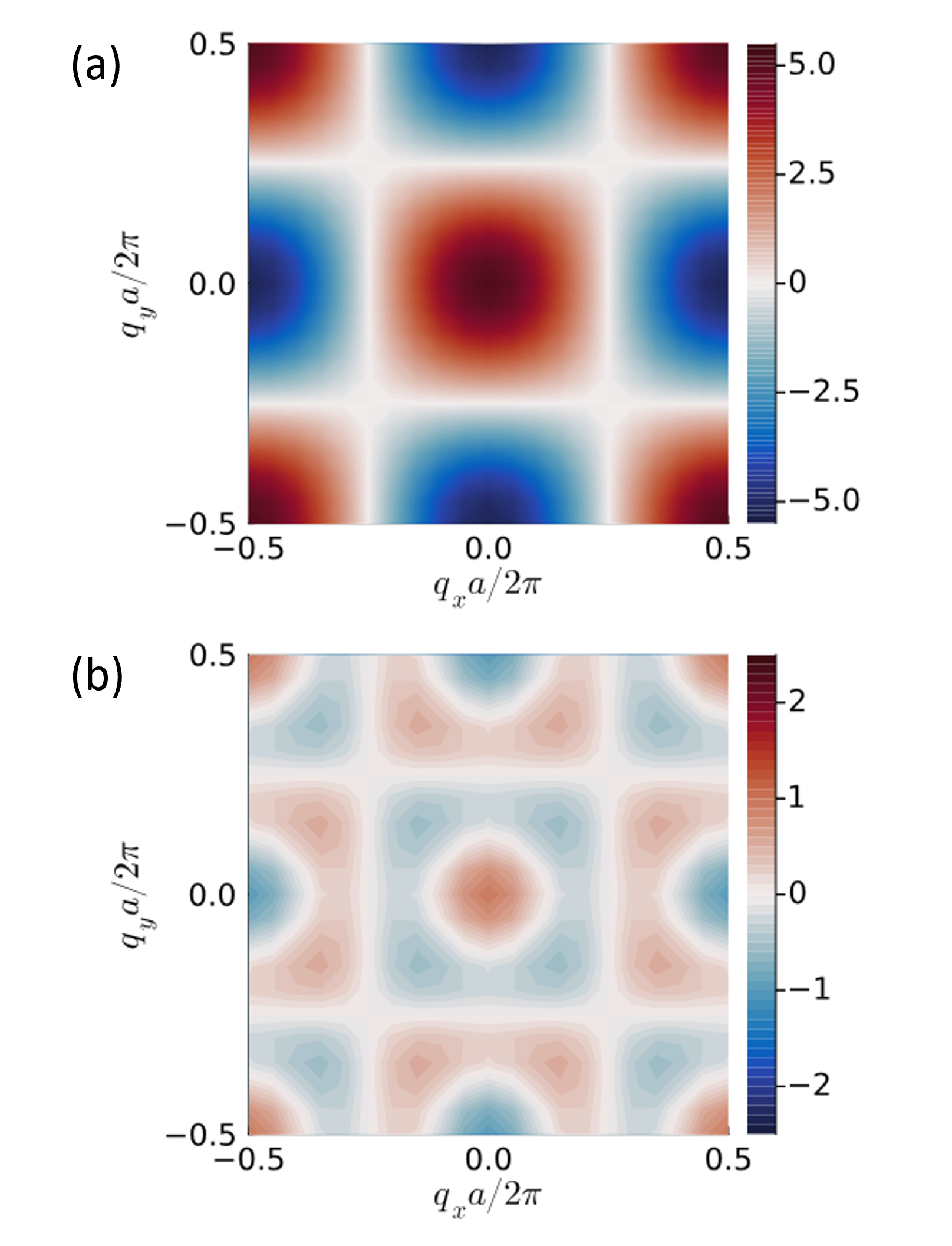}
    \caption{ Contour plots of $\Delta U({\bf q})$ in the first Brillouin zone \textcolor{black}{with one Fe per unit cell,} when (a) only nn As clouds, and (b) both nn+nnn As clouds are included, with no other screening. Other parameters are $\alpha_p=10\AA^3, \Omega=6$eV.}
    \label{fig: contours}    
\end{figure}

To conclude this section, in the static case we find a polarization mediated effective interaction, strongly attractive at short ranges, described by:
\begin{align}
    \Delta \hat{U} = {1\over 2}\sum_{i,j\in As} \Delta U_{|i-j|} \hat{n}_i \hat{n}_j. \label{DU-real}
\end{align}
Our results show that this attraction is even stronger than estimated in  Ref. \onlinecite{george2}, because of the non-negligible contributions of the As ions from second coordination rings. As such, this (over)screening mechanism is  even more likely to play a role in the appearance of superconductivity and other correlated orders. To better visualize the momentum (${\bf q}$) dependence of $\Delta U_{n}$, we inspect its Fourier transform
\begin{align}
    \Delta \hat{U} = {1\over 2N}\sum_{{\bf k, k', q}\atop \sigma,\sigma'} \Delta U({\bf q}) c^\dagger_{{\bf k+q}, \sigma} c^\dagger_{{\bf k'-q}, \sigma'}  c_{{\bf k'}\sigma'} c_{{\bf k}\sigma} , \label{DU-mom}
\end{align}
where
\begin{align}
   \Delta U({\bf q}) = \sum_{i\in As} \Delta U_{|i|} e^{i{\bf q}\vdot {\bf R_i}}. \label{DU-real}
\end{align}
If only the contribution of the nn As is included, then $\Delta U({\bf q}) =\Delta U_0 + 2 \Delta U_1 [\cos(q_xa) + \cos(q_ya)] + 4 \Delta U_2 \cos(q_xa) \cos(q_ya)$. If the nnn As are  also included, then $\Delta U$ contains contributions for up to the 7$^{th}$ nearest Fe-Fe neighbors.  

Fig. \ref{fig: contours} displays $\Delta U({\bf q}) $ for these two cutoffs. As expected, inclusion of the second ring of As clouds results in a more structured pattern (higher harmonics). We reiterate that these results assume no screening when calculating the electric fields. Consideration of screening would suppress the contributions of the nnn As, reducing the relative strength of the longer range interactions. We emphasize that the momentum-structure of $\Delta U$ with local minima at finite-${\bf q}$ may mediate the emergence of (in)commensurate dynamic charge order \cite{Caprara, Boschini}.

%%%%
\section{Results in the time-dependent case}
Now we discuss how an external electric field may affect the polarization of the As ions, thus modulating the effective interactions between carriers (polarons). 

\subsection{Single polaron}

Consider any of the As polarized by the static electric field ${\bf E}$ created by the charge carrier (we continue to include in the calculation only nn and nnn As). Generically, the laser field ${\bf E}_a$ will have a component parallel to ${\bf E}$ and a component perpendicular to it. Indeed, as a direct consequence of the geometrical arrangement of As around the Fe which hosts the charge, it is not possible to choose a laser field polarization ${\hat \epsilon}_a$ so that it is simultaneously parallel (or perpendicular) to all ${\bf E}$ at all polarized As sites. If ${\bf E_a}$ is parallel to ${\bf E}$, it excites holes from their static ground-state $E_0(g)$ into the highest eigenstate  $E_1(g)$. Instead, if ${\bf E_a} \perp {\bf E}$, it excites holes from  $E_0(g)$ into a $p_{\perp}$ orbital with energy $\Omega$. Thus, even though the resonance is at $\Omega$ for the unpolarized As, the resonant frequencies for a polarized As are at $\omega_{res,1}(g)=E_1(g) - E_0(g)= \sqrt{\Omega^2+4g^2}$ and $\omega_{res,2}(g)=\Omega - E_0(g) = [\Omega + \sqrt{\Omega^2+4g^2}]/2$. If we include both coordination rings with their corresponding couplings $g_1$ and $g_2$, we expect to observe a rich beating pattern in the presence of a laser field with $\omega_a \approx \Omega$.

We define the time-dependent average polaron energy $E_P(t)$ as the change in the average energy of the system due to the presence of a carrier on an Fe. As such, to find $E_P(t)$ in the presence of the laser field, we need to sum up contributions from the various As polarized by this charge, and subtract the energy of the same As sites in the absence of the charge carrier (the latter is no longer zero, unlike in the static case). This requires us to solve the time-dependent Schr\"odinger's equation for each As; the details are discussed in Appendix A.

\begin{figure}[!t]
    \centering
    \includegraphics[width=\columnwidth]{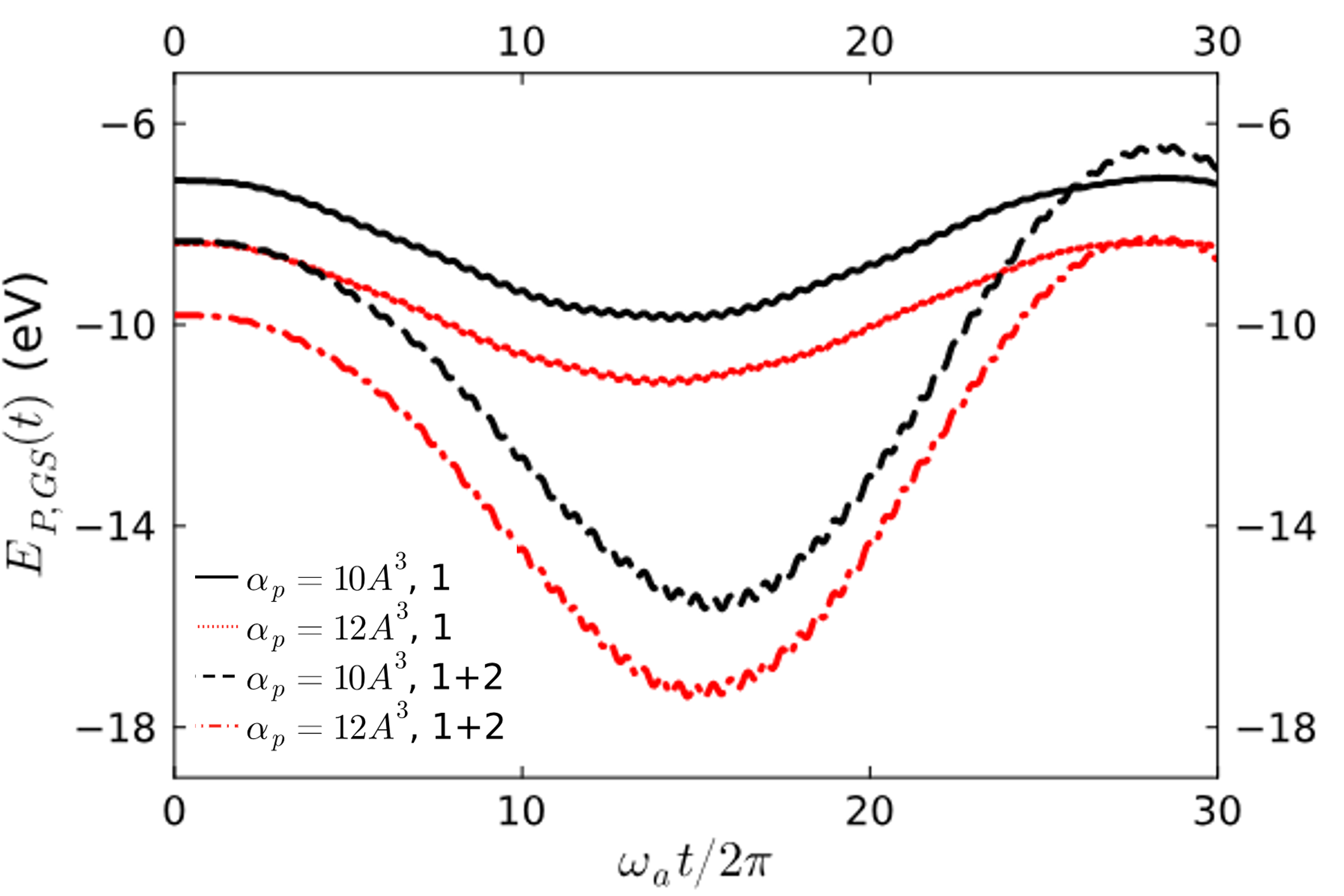}
    \caption{Energy of a single polaron under laser perturbation with polarization ${\hat \epsilon}_a\parallel {\bf {\hat e}}_1= (\hat{x}+\hat{y})/\sqrt{2}$. Here we assume the s-p transition at typical value of $\Omega = 6{eV}$, and simulate laser perturbation with frequency $\omega_a = 5.8{eV}$ and field strength $5{MV/cm}$.}
    \label{fig: laser_Egs}
\end{figure}

% Figure after laser perturbation
\begin{figure*}
    \centering
    \includegraphics[width=\linewidth]{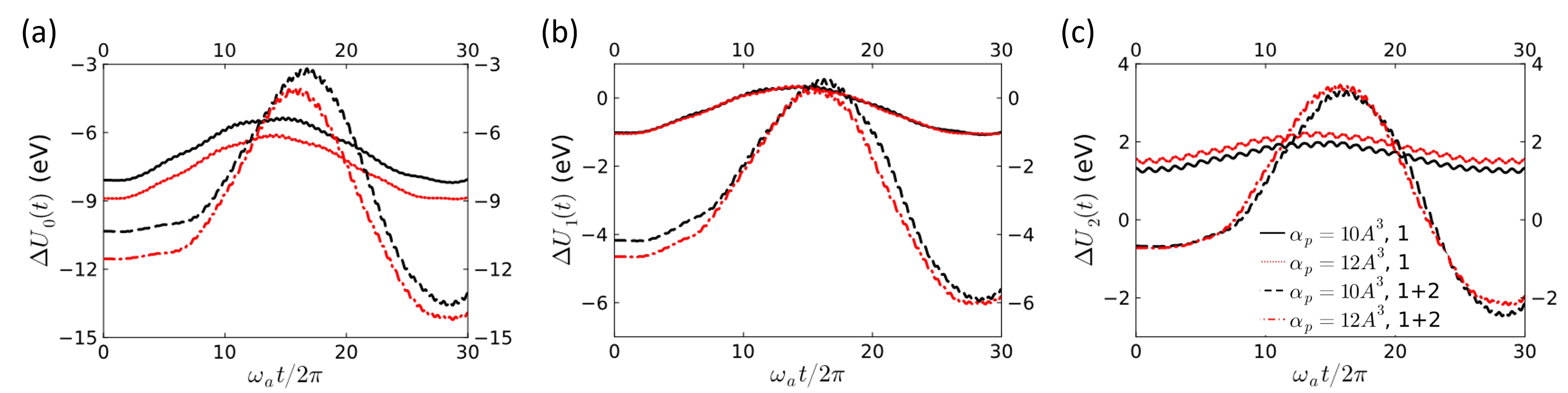}
    \caption{Effective attraction between (a) on-site, (b) nearest-neighbour, and (c) second-nearest-neighbour polarons under laser excitation as a function of time. We assume a sharp s-p transition at $\Omega = 6eV$. The simulated laser has frequency $\omega_a={5.8}{eV}$ and is oriented along $\vec{e_1}$. The nn charges are aligned parallel to $\vec{e_1}$ and the nnn charges are placed parallel to $\vec{x}$.}
    \label{fig: laser_effective_attractions}    
\end{figure*}

Typical results for $E_P(t)$ in the presence of a laser with polarization ${\hat \epsilon}_a\parallel {\bf {\hat e}}_1= (\hat{x}+\hat{y})/\sqrt{2}$ are shown in Fig. \ref{fig: laser_Egs}.  We chose the laser frequency $\omega_a={5.8}{eV}$ to be close to, but not resonant to, the $s$-$p$ transition energy of $\Omega={6}{eV}$ to avoid on-resonance response of the unpolarized As sites, which would overwhelm the polaronic-type contributions from the polarized (but off-resonant) sites.  All the curves in Fig. \ref{fig: laser_Egs} show small-amplitude forced oscillations with frequency $\omega_a$, superimposed over a much higher amplitude beat at a frequency around $\Omega - \omega_a\approx  \omega_a / 30$ due to the unpolarized sites. When both coordination rings are included there are additional beat frequencies, coming from the least polarized As, whose resonance frequencies $\omega_{res,1/2}$  are closest to $\Omega$.

\subsection{Effective interactions between two polarons}

To calculate the light-induced transient evolution of the effective interaction $\Delta U_n(t)= E_n(t) - E_{\infty}(t) $ for two carriers that are $n^{th}$ nearest neighbors, we mirror the static calculation and calculate $E_{\infty}(t) = 2 E_P(t)$ of two polarons located very far apart.  The calculation of $E_n(t)$ is similar to that of $E_P(t)$ displayed in Fig. \ref{fig: laser_Egs}, but involving more values of the couplings  (and hence more resonances) depending on the positioning of various As with respect to the Fe sites hosting the two charges.

% Contour plot after laser perturbation
\begin{figure*}
    \centering
    \includegraphics[width=0.8\linewidth]{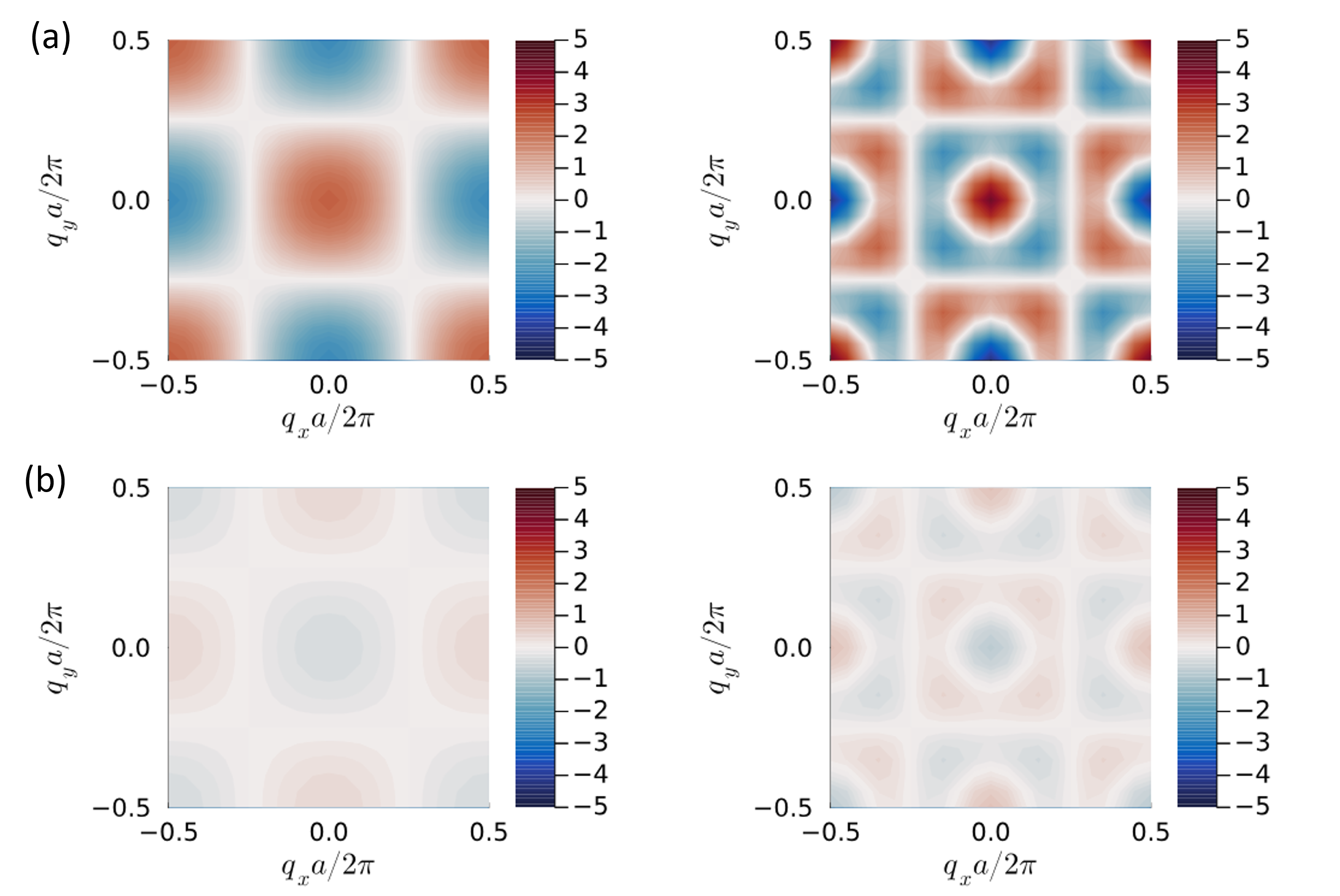}
    \caption{Contour plots of the differential $\Delta U({\bf q},t) - \Delta U({\bf q},0)$  \textcolor{black}{in the first Brillouin zone} at (a) maximum repulsion $\omega_a t/2\pi \approx 17$, (b) maximum attraction $\omega_a t/2\pi \approx 27.5$.  The reference point of $\Delta U({\bf q},0)$ are identical to the static results in Figure \ref{fig: contours}. The model includes As ions in ring 1 (left panels) and ring 1+2 (right panels). The  laser electric field has amplitude of {5}{MV/cm}, frequency of ${5.8}${eV}, and is polarized along $\vec{e}_1$. }
    \label{fig: laser_effective_attractions_ftx}    
\end{figure*}

% Results for ring 1 and ring 2 effective attractions
Figure \ref{fig: laser_effective_attractions} shows the time dependence of the effective attraction between on-site,  nearest- and second-nearest-neighbour polarons upon a light excitation with ${\hat \epsilon}_a\parallel {\bf {\hat e}}_1$. Without loss of generality, we fix the position of the charges in these plots so that nn charges are along  $\hat{x}$  and nnn charges are along ${\bf {\hat e}}_1$. We report a time-dependence behavior mirroring that of a single polaron in Fig. \ref{fig: laser_Egs}. Interestingly, we see that the short-range effective interactions turn more repulsive (even when they started strongly attractive) at around $T_m \sim \pi/(\Omega-\omega_a)$. This is reasonable considering that $T_m$ roughly defines when the external field induces its maximum polarization of As ions. This light-induced polarization competes with that induced by the static electric field of the charge carriers, and therefore affects the screening of their effective interactions.
In particular, the stronger stability of individual polarons for $\omega_a T_m /2\pi \sim 17$ translates in an effectively more repulsive (less screened) effective interaction between carriers at those times.  The amplitude of the variations of the various $\Delta U_n(t)$ terms is very considerable even for a laser electric field of 5 MV/cm, for which the corresponding $g_a= 0.05$eV is two orders of magnitude smaller than $g_1\approx 2.5$eV. 

%and nnn charges are always along $\vec{x}$, and move around the laser in different directions. Later we investigate the effect of changing direction of the laser relative to charges in fig \ref{fig:laser}, which suggests the relative direction of the laser has minimal effect on the response.

The variations of the various $\Delta U_n(t)$ combine into an overall variation of their Fourier transform $\Delta U({\bf q}, t)$. The differences $\Delta U({\bf q}, t) - \Delta U({\bf q}, 0)$ are displayed in Figure \ref{fig: laser_effective_attractions_ftx} for $\omega_a t/2\pi \approx 17$ and for $\omega_a t/2\pi \approx 27.5$. 
The significant changes with time suggest that the effects of the laser should be readily visible from modulations of the properties of the electronic state of the system, as well as underlying collective modes.

\begin{figure}
    \includegraphics[width=0.85\columnwidth]{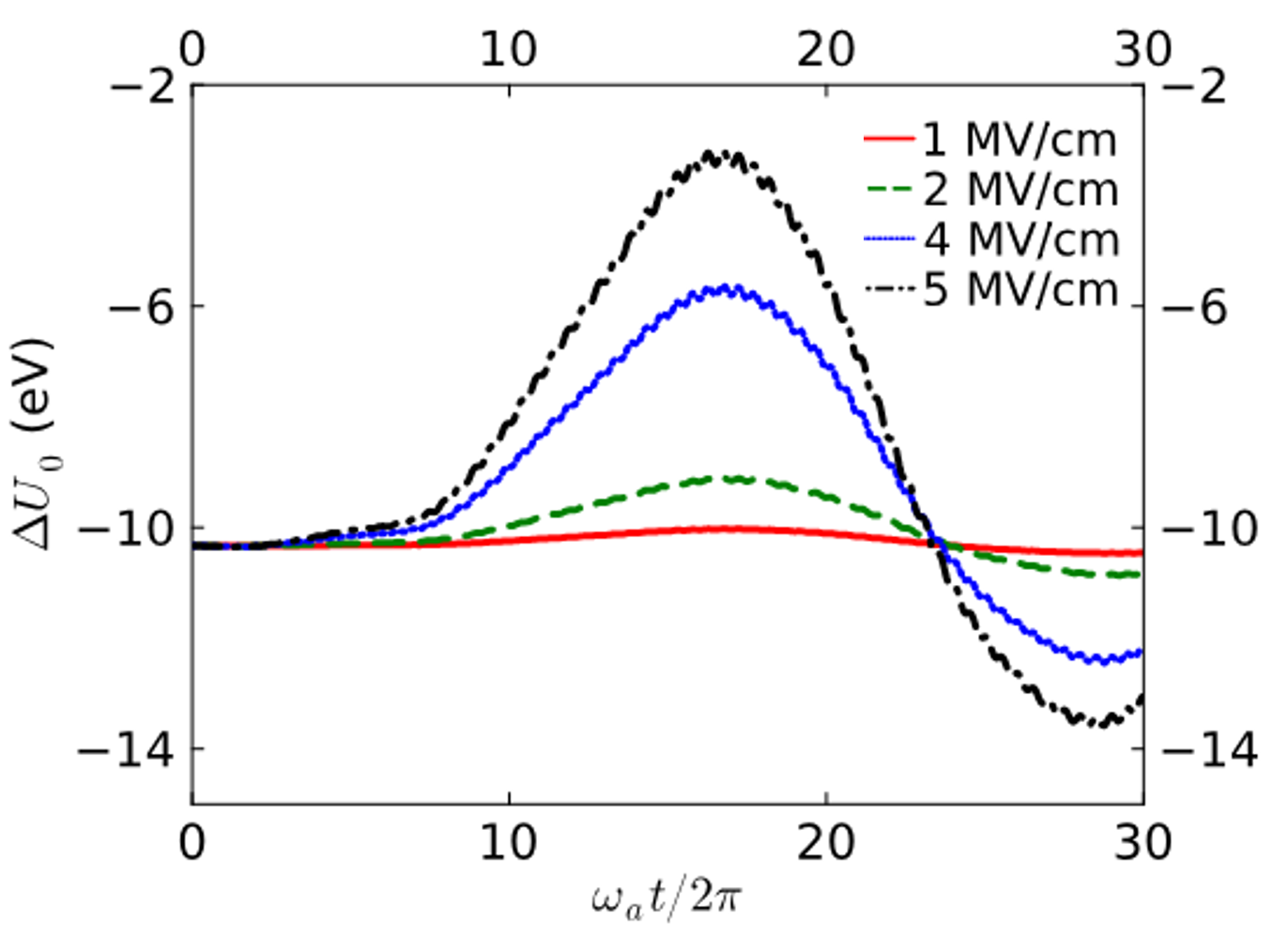}
    \includegraphics[width=0.85\columnwidth]{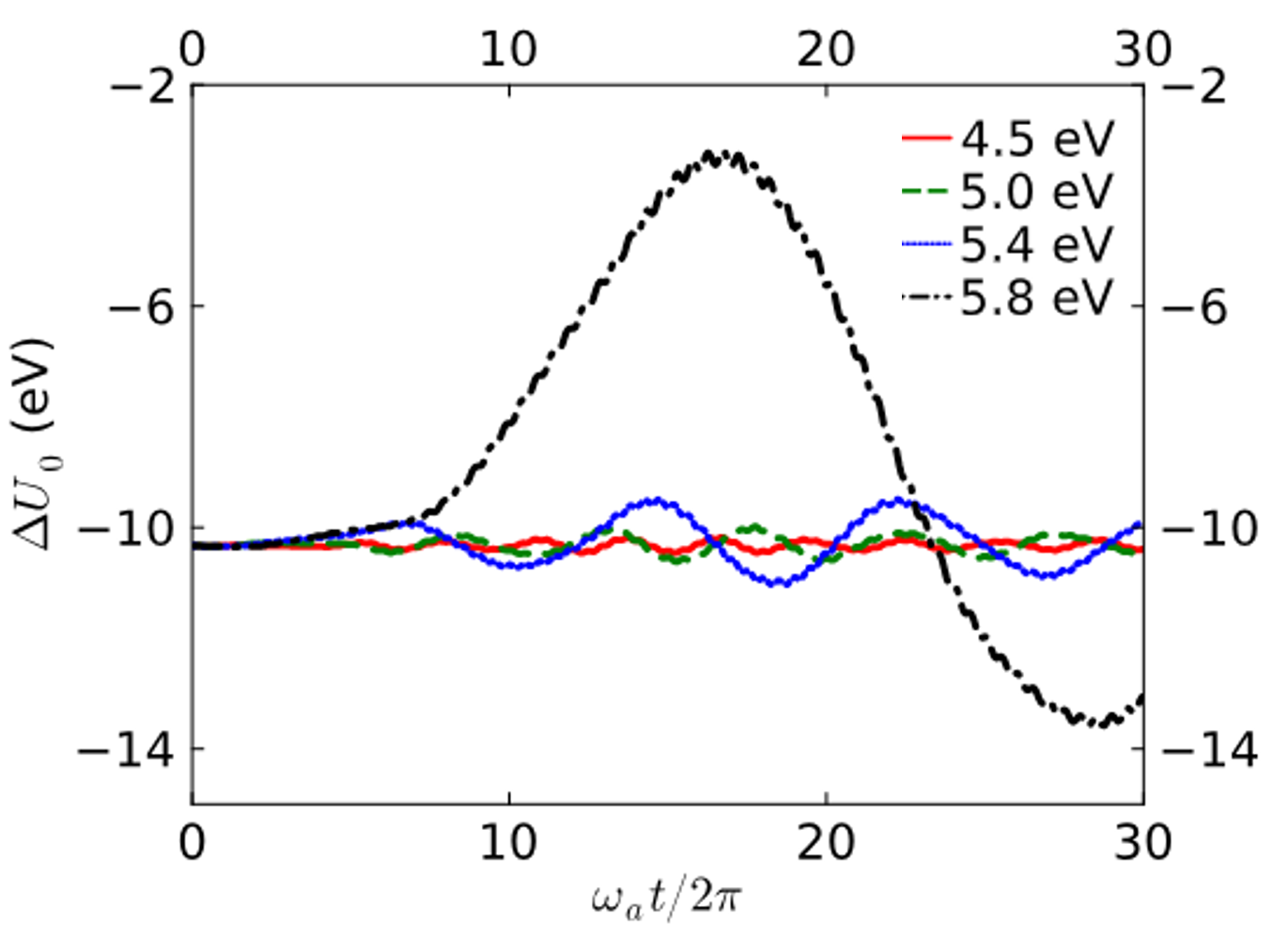}
    \includegraphics[width=0.85\columnwidth]{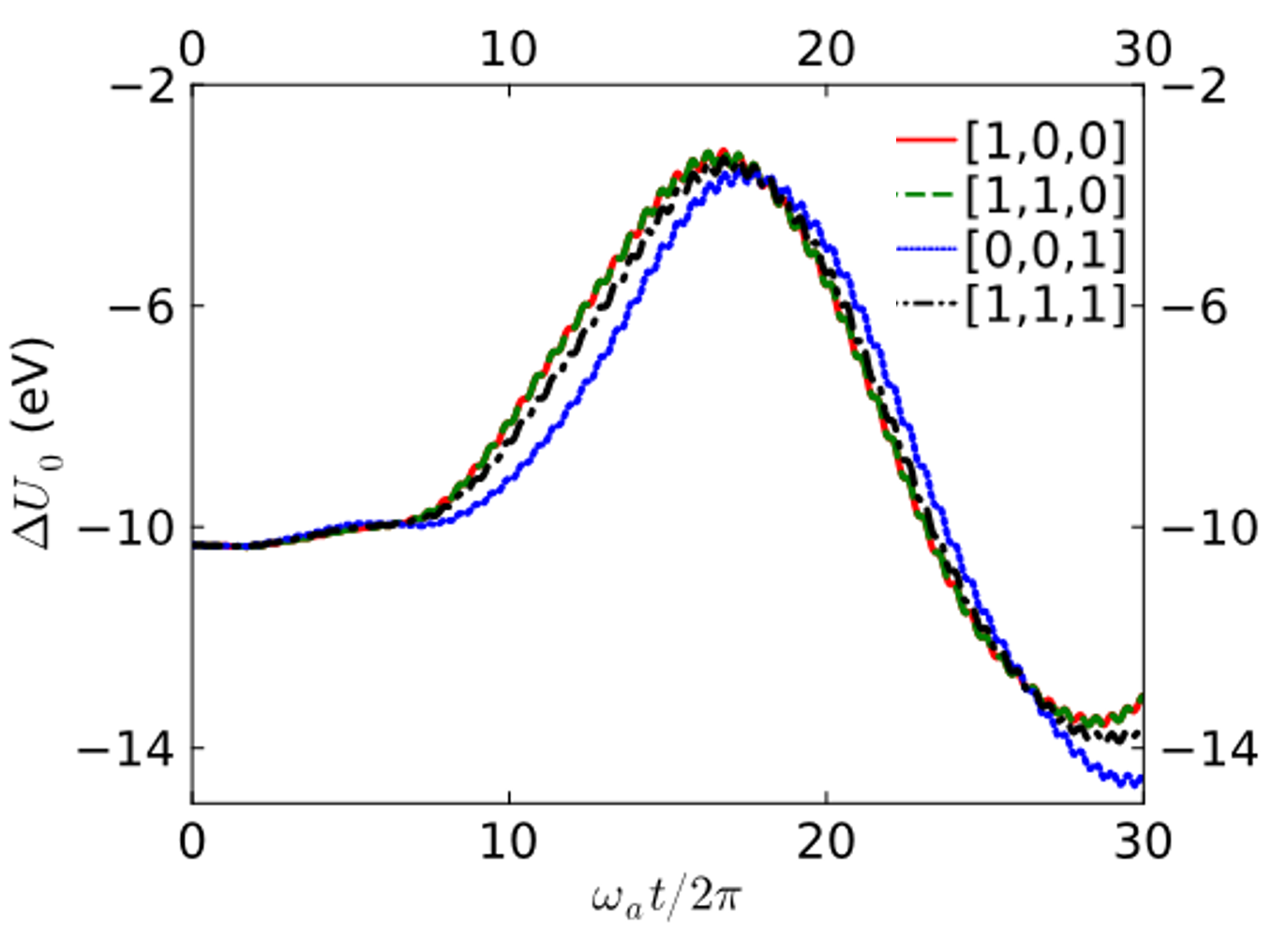}
    
    \caption{Effective attraction between on-site polarons under laser excitations with different (a) laser field amplitudes, (b) laser excitation frequencies and (c) laser polarization directions indicated by coordinates of $\vec{e}_1, \vec{e}_2,\vec{e}_3$. In all these simulations the nn charges are along $\vec{e}_1$, and the results include contributions from their nn+nnn As sites. Unless otherwise labelled on the legend, the simulated laser electric field is ${5}{MV/cm}$, laser frequency is ${5.8}{eV}$, and it is polarized along $\vec{e}_1$. }
    \label{fig: laser_attr}
\end{figure}

Figure \ref{fig: laser_attr} explores the dependence of $\Delta U_0(t)$ on the intensity (top panel), frequency (middle panel)  and polarization (bottom panel) of the laser. We only show results for $\Delta U_0(t)$ because the other effective interactions display analogous responses, though with smaller amplitudes. As expected, the amplitude of the response increases monotonically with  increasing laser intensity (top panel) and with decreasing detuning $\Omega - \omega_a$ (middle panel).

% Results for different laser parameters

\begin{figure}
    \centering
    \includegraphics[width=0.9\columnwidth]{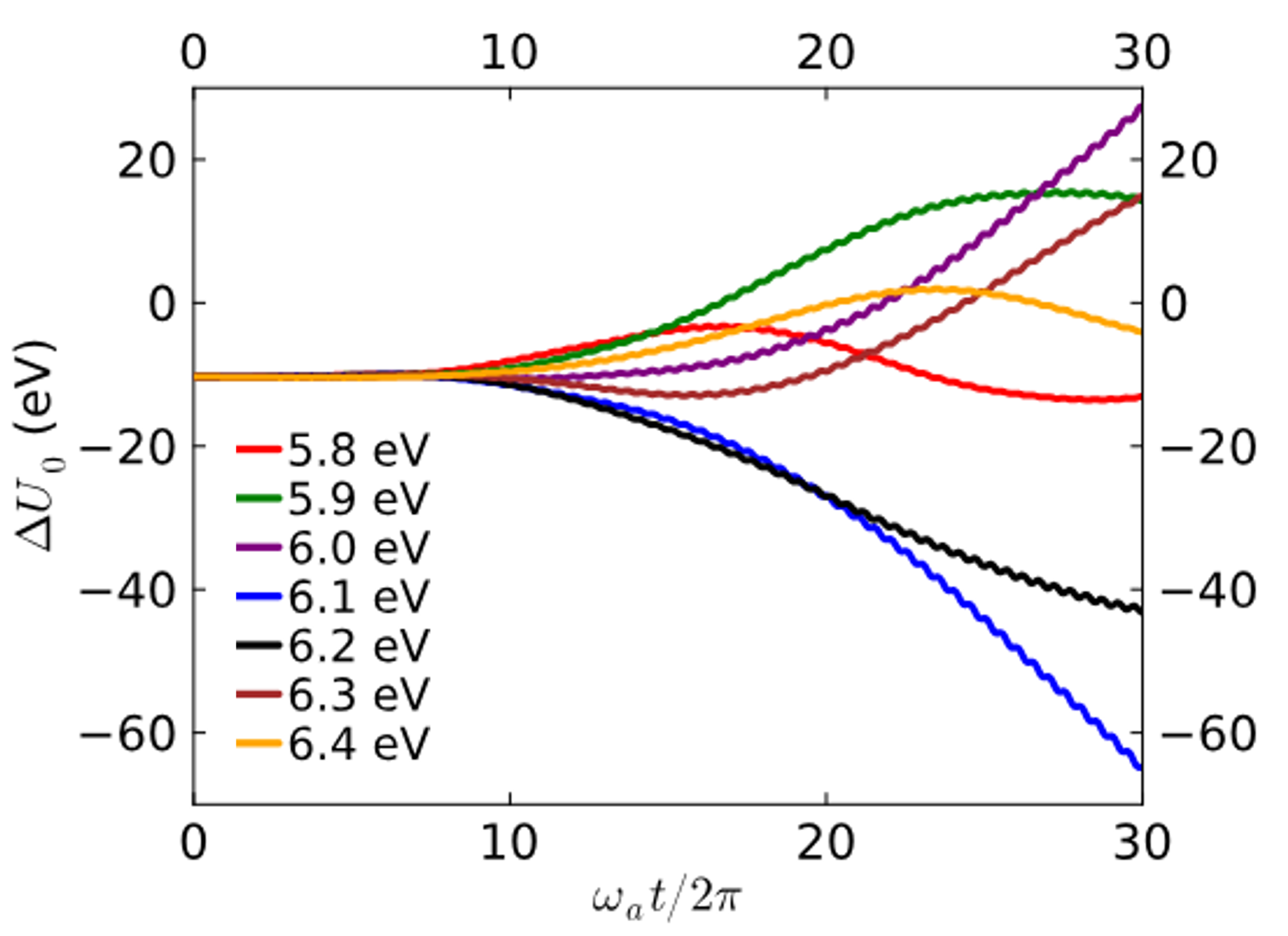}
    \caption{Same as in Fig. \ref{fig: laser_attr}(b), but for laser frequencies between $5.8eV$ and $6.4eV$.}
    \label{fig: laser_high_freqs}
\end{figure}

The bottom panel of Fig. \ref{fig: laser_attr} shows that the polarization of the laser field does not play much of a role in the amplitude of the response. This is because of the geometry of the lattice, which leads to static electric fields at various polarized As sites being 'hedgehog'-like. This is a key result because it should allow one to isolate changes in the behavior of the system due to the mechanism discussed here  from other possible changes induced by ${\bf E}_a$. For instance, apart from polarizing the As, an external electric field could also excite electrons from below to above the Fermi energy, if the band structure is such that direct electronic transitions with energy $\omega_a$ are possible. Such electronic excitations should, however, be very sensitive to the laser polarization, which would have to be aligned with the local dipole of that transition. 

All the results discussed so far are for a laser frequency $\omega_a < \Omega$, meaning that the (otherwise) unpolarized As orbitals are closest to resonance with the laser and determine the period of the beats. Figure \ref{fig: laser_high_freqs} mirrors panel (b) of Fig. \ref{fig: laser_attr} but for frequencies $\omega_a > \Omega$, where some of the polarized orbitals can become resonant. Indeed, beside the expected strong resonance at $\omega_a = \Omega = 6$eV, we accidentally hit close to another resonance for $\omega_a = 6.1$eV, which could be due to alignment of $\omega_a$ with either $\omega_{res,1}$ or $\omega_{res,2}$ of some of the polarized ions. 

Of course, in a more realistic description of the system these resonances would be tempered by a variety of ways in which the polarized As ions could dissipate their energy into other subsystems, like the lattice and other electronic degrees of freedom. As such, these results are not meant to be quantitatively accurate, instead they only serve as a proxy for illustrating the non-trivial influence of an external electric field on the polarization of As ions, and through that, onto the polarization-controlled screening of carrier-carrier interactions in the system.

\section{Discussions}

In this study, we investigated whether an external field resonant to the As 5s-4p transition, with a coupling magnitude nearly two orders of magnitude lower than the coupling of the As to the field caused by a nearby charge, can be an effective means to drive a sizable modulation of the effective polarization-mediated interaction between carriers in FeAs.

Our simulations suggest that the effect can be very considerable, especially if the external field happens to be close to resonant with some of the many possible resonances of the polarized As ions. The observed modulation of the effective carrier-carrier interactions by a few eV, and the change of some terms from attractive into repulsive, should have a very considerable effect on the electronic properties of FeAs. This extends beyond superconductivity, and include all other possible ordered states (magnetic, charge order, etc). The significant modulation of the effective interactions driven by the external electric field could either switch the system between different ordered states or between an ordered state and a disordered one. Even if the system is deep inside an ordered state and the modulation is not sufficiently strong to drive a phase transition, the characteristic energy scales associated with the collective excitations of that ordered state should change in response to the applied external electric field. 

While here we focused on  the simplest scenario of applying a continuous-wave excitation, pump-probe spectroscopic approaches may also provide direct experimental evidence of light-induced modulation of the electronic polarizability in FeAs and related compounds. Our simulations predict a significant modification of the interactions within a few tens of optical cycles, corresponding to approximately 20 fs for photon energies in the 6 eV range. Pump-probe table-top all-optical spectroscopies \cite{N1} or free-electron-laser-based X-ray scattering approaches\cite{baykusheva,mitrano2020probing}  can achieve these temporal resolutions and extract light-induced modulations of the Coulomb interaction. The effect of an ultrafast pulse on the effective carrier-carrier interactions can be calculated with the same methods we used, by changing the profile of the external field in Eq. \ref{eq: H_laser}.

However, in order to make more quantitative predictions, it is imperative to first relax the many approximations we made in this model. In particular, the  Fe-superconductors are multi-band systems \cite{MM1,MM2,MM3} with non-trivial electronic bands, instead of the flat (dispersionless) single orbital case studied here. \textcolor{black}{It was shown in previous work that the dressing by As polarization clouds results in a sizable renormalization of these bands,\cite{george2} roughly by a factor of 2. It remains for future work to determine whether and how an external laser field would affect this renormalization of the electronic band structure, in addition to the modulation of the effective carrier-carrier interactions discussed here.  We expect that this renormalization is both direct (if the laser modulates the polarization cloud of a quasiparticle, that will affect its effective hopping between sites) but also indirect (even within a Hartree-Fock approximation, a modulation of the effective interactions will induce a change of the single quasiparticle dispersion). If this band renormalization turns out to be significant, time- and angle-resolved photoemission spectroscopy could be employed to track light-induced band structure renormalization over the entire momentum space \cite{ARPES_review}.}

Another \textcolor{black}{very} necessary improvement is to allow the As $s$ and $p$ levels to also broaden into bands. Our simple calculation presented here assumes that each As is instantaneously polarized by the various fields (from charges or from the laser), and these polarization clouds do not have any dynamics of their own. In reality, the finite bandwidth of these bands means that the electron-hole excitation created by the electric fields can hop onto neighboring As sites, thus affecting the nature of the polarization clouds and, therefore, their overall effect on both quasiparticles' dispersion and their effective interactions. \textcolor{black}{Potentially more damaging for the proposal made here is the fact that a finite bandwidth of the As levels may turn the discrete eigenstates sketched in Fig. 1(c) into broad resonances inside these bands. If this were to happen, then the large effects illustrated in Fig. 9, obtained when the laser frequency is resonant with some of these discrete levels, would likely be washed out. While this possibility obviously needs to be investigated carefully, we expect that at least some of these discrete states are located outside the bandwidths of their respective bands, and therefore survive as discrete levels even when the hybridization is turned on. For instance, consider an As adjacent to an Fe hosting two charge carriers. Their electric field lowers the ground-state of the As from energy '0' (in fact, the center of the $5s$-band if hopping between As states is allowed) to ${1\over 2}[\Omega - \sqrt{\Omega^2+ 16 g^2}]\approx -2.8$eV for the typical values we used here. This level will be located below the $5s$ band (and thus remain discrete and exhibit a resonant effect) if the bandwidth of the latter is less that $\approx 5.6$eV, which seems likely considering the large distance between neighbor As given their alternations above/below the Fe plane.  }

\textcolor{black}{Finally, our model also ignores the effects of the doping ions that donated the charge carriers to the FeAs layer. This is a reasonable approximation if these ions are located sufficiently far enough so that their electric fields within the FeAs layer are much smaller than those of the carriers, but this assumption needs to be tested. }

This is why we believe that before we can proceed to suggest the most promising experimental approaches to observe and untangle the effects of the As polarization on the behavior of the Fe superconductors, all these above mentioned approximations need to be relaxed. Understanding the details of all this phenomenology is not a trivial exercise and we leave it to future work, \textcolor{black}{which might use methods like  those employed in Ref. \onlinecite{Janez} for a somewhat related but still very different problem}. Nevertheless, we believe that the work presented here provides a strong motivation for engaging in these much more complicated calculations: the effects we find are so significant that it seems very likely that they will persist in more detailed models, albeit with quantitative and perhaps even qualitative changes.

\begin{acknowledgments}We thank George Sawatzky, Andrea Damascelli and  Matteo Mitrano for insightful discussions.  We acknowledge support from the Natural Sciences and Engineering Research Council of Canada, the Stewart Blusson Quantum Matter Institute, and the Max-Planck-UBC-UTokyo Center for Quantum Materials. We  also acknowledge support from the Fonds de recherche du Québec - Nature et Technologie (FRQNT) and the Ministère de l'Économie, de l'Innovation et de l'Énergie - Québec.
\end{acknowledgments}

%% Appendix Calculation
\appendix
\section{Calculating the time-dependent average polarization energy for an As ion}

Here we describe briefly the calculation of the time-dependent average energy contributed by one As ion subject to the laser field. The results shown in Section IV include contributions from all the polarized As ions. 

For simplicity, we drop the site and spin index and denote the hole operators of the As of interest by $s, p_\lambda$, $\lambda=1,2,3$. Let ${\bf E}$ be the {\em static} electric field at the As site, due to the various charge carriers present in the system,  $g= \sqrt{\alpha_p\Omega} E/2$ be its corresponding coupling, $A_\lambda = {\bf E} \cdot {\bf \hat{e}}_\lambda/ E$ define the orientation of the electric field along various axes, and $\bar{p} = \sum_\lambda A_\lambda p_\lambda$ be the $p$ orbital oriented parallel to the static electric field.

The static part of the Hamiltonian $${\hat h}_{As}+ {\hat h}_{p}= \Omega \sum_{\lambda} p^\dagger_\lambda p_\lambda + g (s^\dagger \bar{p} + h.c.) = \sum_{a=0}^3 E_a \gamma_a^\dagger \gamma_a$$   is diagonalized by the operators $\gamma_{0}^\dagger = \cos\alpha s^\dagger - \sin\alpha {\bar p}^\dagger$, $\gamma_1^\dagger = \sin \alpha s^\dagger + \cos\alpha {\bar p}^\dagger$, $\gamma_2^\dagger = ( A_3 p_2^\dagger - A_2 p_3^\dagger )/\sqrt{A_3^2 + A_2^2}$ and $\gamma_3^\dagger =  [- (A_2^2 + A_3^2) p_1^\dagger + A_1 A_2 p_2^\dagger + A_1 A_3  p_3^\dagger ]/\sqrt{A_2^2 + A_3^2}$ where $E_0 = \frac{1}{2} ( \Omega - \sqrt{\Omega^2 + 4 g^2})$, $E_1 = \frac{1}{2} ( \Omega + \sqrt{\Omega^2 + 4 g^2})$, $E_2=E_3=\Omega$, and $\cos \alpha = \sqrt{(1 + \Omega/\sqrt{\Omega^2+4g^2})/2}$.

Because they are degenerate, the operators $\gamma_2, \gamma_3$ can be chosen to be along any two directions orthogonal to ${\bf E}$. The choice made above is optimal when the laser field is assumed to be parallel to  ${\bf \hat{e}}_1={\bf \epsilon}_a$, so that $\gamma_2$ is orthogonal to both ${\bf E}$ and ${\bf \epsilon}_a$. For a different ${\bf \epsilon}_a$, $\gamma_2$ is rotated accordingly.

With this choice, $\hat{h}_{ext}(t)= g_a \cos(\omega_a t) (s^\dagger p_1 + h.c.)$ can be recast in terms of only $\gamma_0, \gamma_1$ and $\gamma_3$, namely:
\begin{align}
\hat{h}_{ext}  &= g_a(t)  \Psi^\dagger \begin{bmatrix}
        \sin(2\alpha) & -\cos(2\alpha) & \cos\alpha A_{\perp} \\
        -\cos(2\alpha) & -\sin(2\alpha) & \sin\alpha A_{\perp}\\
        \cos\alpha E_{\perp}& \sin\alpha A_{\perp} & 0
    \end{bmatrix} \Psi
    \label{H1}
\end{align}
where $g_a(t) \equiv g_a \cos(\omega_a t)$ and $\Psi^\dagger = \begin{bmatrix}\gamma_0^\dagger & \gamma_1^\dagger & \gamma_3^\dagger \end{bmatrix}$. $E_{\perp} = \sqrt{E_2^2 + E_3^2}$ and $ A_{\perp} = \sqrt{A_2^2 + A_3^2}$ are the components perpendicular to the laser polarization ${\hat \epsilon}_a$.

To solve Schr\"odinger's equation $i {\partial \over \partial t} |\phi(t)\rangle = {\hat h}_{tot}(t) |\phi(t)\rangle $ we write $|\phi(t)\rangle = \sum_{a\ne 2}^{}d_a(t) e^{-i E_a t} \gamma_a^\dagger|0\rangle$, leading to a system of three coupled differential equations: $\dot{d}_0 = -i g_a(t) [\sin(2\alpha)  d_0 - \cos(2\alpha) e^{-(E_1 - E_0) t} d_1 + \cos\alpha A_{\perp} e^{-(E_3 - E_0) t} d_3] $ and similarly for $\dot{d}_1$ and $\dot{d}_3$. We integrate these equations numerically using  the fourth order Runge-Kutta method, starting from the ground-state  $|\phi(0)\rangle = \gamma^\dag_0|0\rangle$, {\em i.e.}  $d_0(0)=1, d_1(0)=d_3(0)=0$.

The time-dependent average polarization energy of this As ion is then found using $E(t) = \langle \phi(t) | \hat{h}_{tot}(t)| \phi(t)\rangle$.

\bibliographystyle{apsrev4-1}
\bibliography{references}

\begin{thebibliography}{30}%
\makeatletter
\providecommand \@ifxundefined [1]{%
 \@ifx{#1\undefined}
}%
\providecommand \@ifnum [1]{%
 \ifnum #1\expandafter \@firstoftwo
 \else \expandafter \@secondoftwo
 \fi
}%
\providecommand \@ifx [1]{%
 \ifx #1\expandafter \@firstoftwo
 \else \expandafter \@secondoftwo
 \fi
}%
\providecommand \natexlab [1]{#1}%
\providecommand \enquote  [1]{``#1''}%
\providecommand \bibnamefont  [1]{#1}%
\providecommand \bibfnamefont [1]{#1}%
\providecommand \citenamefont [1]{#1}%
\providecommand \href@noop [0]{\@secondoftwo}%
\providecommand \href [0]{\begingroup \@sanitize@url \@href}%
\providecommand \@href[1]{\@@startlink{#1}\@@href}%
\providecommand \@@href[1]{\endgroup#1\@@endlink}%
\providecommand \@sanitize@url [0]{\catcode `\\12\catcode `\$12\catcode `\&12\catcode `\#12\catcode `\^12\catcode `\_12\catcode `\%12\relax}%
\providecommand \@@startlink[1]{}%
\providecommand \@@endlink[0]{}%
\providecommand \url  [0]{\begingroup\@sanitize@url \@url }%
\providecommand \@url [1]{\endgroup\@href {#1}{\urlprefix }}%
\providecommand \urlprefix  [0]{URL }%
\providecommand \Eprint [0]{\href }%
\providecommand \doibase [0]{http://dx.doi.org/}%
\providecommand \selectlanguage [0]{\@gobble}%
\providecommand \bibinfo  [0]{\@secondoftwo}%
\providecommand \bibfield  [0]{\@secondoftwo}%
\providecommand \translation [1]{[#1]}%
\providecommand \BibitemOpen [0]{}%
\providecommand \bibitemStop [0]{}%
\providecommand \bibitemNoStop [0]{.\EOS\space}%
\providecommand \EOS [0]{\spacefactor3000\relax}%
\providecommand \BibitemShut  [1]{\csname bibitem#1\endcsname}%
\let\auto@bib@innerbib\@empty
%</preamble>
\bibitem [{\citenamefont {Bardeen}\ \emph {et~al.}(1957)\citenamefont {Bardeen}, \citenamefont {Cooper},\ and\ \citenamefont {Schrieffer}}]{BCS}%
  \BibitemOpen
  \bibfield  {author} {\bibinfo {author} {\bibfnamefont {J.}~\bibnamefont {Bardeen}}, \bibinfo {author} {\bibfnamefont {L.~N.}\ \bibnamefont {Cooper}}, \ and\ \bibinfo {author} {\bibfnamefont {J.~R.}\ \bibnamefont {Schrieffer}},\ }\href {\doibase 10.1103/PhysRev.108.1175} {\bibfield  {journal} {\bibinfo  {journal} {Phys. Rev.}\ }\textbf {\bibinfo {volume} {108}},\ \bibinfo {pages} {1175} (\bibinfo {year} {1957})}\BibitemShut {NoStop}%
\bibitem [{\citenamefont {Alexandrov}\ and\ \citenamefont {Mott}(1995)}]{Alexandrov}%
  \BibitemOpen
  \bibfield  {author} {\bibinfo {author} {\bibfnamefont {A.~S.}\ \bibnamefont {Alexandrov}}\ and\ \bibinfo {author} {\bibfnamefont {N.~F.}\ \bibnamefont {Mott}},\ }\href {\doibase 10.1142/2784} {\emph {\bibinfo {title} {Polarons and Bipolarons}}}\ (\bibinfo  {publisher} {World Scientific, Singapore},\ \bibinfo {year} {1995})\BibitemShut {NoStop}%
\bibitem [{\citenamefont {Zhang}\ \emph {et~al.}(2023)\citenamefont {Zhang}, \citenamefont {Sous}, \citenamefont {Reichman}, \citenamefont {Berciu}, \citenamefont {Millis}, \citenamefont {Prokof'ev},\ and\ \citenamefont {Svistunov}}]{John1}%
  \BibitemOpen
  \bibfield  {author} {\bibinfo {author} {\bibfnamefont {C.}~\bibnamefont {Zhang}}, \bibinfo {author} {\bibfnamefont {J.}~\bibnamefont {Sous}}, \bibinfo {author} {\bibfnamefont {D.~R.}\ \bibnamefont {Reichman}}, \bibinfo {author} {\bibfnamefont {M.}~\bibnamefont {Berciu}}, \bibinfo {author} {\bibfnamefont {A.~J.}\ \bibnamefont {Millis}}, \bibinfo {author} {\bibfnamefont {N.~V.}\ \bibnamefont {Prokof'ev}}, \ and\ \bibinfo {author} {\bibfnamefont {B.~V.}\ \bibnamefont {Svistunov}},\ }\href {\doibase 10.1103/PhysRevX.13.011010} {\bibfield  {journal} {\bibinfo  {journal} {Phys. Rev. X}\ }\textbf {\bibinfo {volume} {13}},\ \bibinfo {pages} {011010} (\bibinfo {year} {2023})}\BibitemShut {NoStop}%
\bibitem [{\citenamefont {Little}(1964)}]{Little}%
  \BibitemOpen
  \bibfield  {author} {\bibinfo {author} {\bibfnamefont {W.~A.}\ \bibnamefont {Little}},\ }\href {\doibase 10.1103/PhysRev.134.A1416} {\bibfield  {journal} {\bibinfo  {journal} {Phys. Rev.}\ }\textbf {\bibinfo {volume} {134}},\ \bibinfo {pages} {A1416} (\bibinfo {year} {1964})}\BibitemShut {NoStop}%
\bibitem [{\citenamefont {Allender}\ \emph {et~al.}(1973)\citenamefont {Allender}, \citenamefont {Bray},\ and\ \citenamefont {Bardeen}}]{ABB}%
  \BibitemOpen
  \bibfield  {author} {\bibinfo {author} {\bibfnamefont {D.}~\bibnamefont {Allender}}, \bibinfo {author} {\bibfnamefont {J.}~\bibnamefont {Bray}}, \ and\ \bibinfo {author} {\bibfnamefont {J.}~\bibnamefont {Bardeen}},\ }\href {\doibase 10.1103/PhysRevB.7.1020} {\bibfield  {journal} {\bibinfo  {journal} {Phys. Rev. B}\ }\textbf {\bibinfo {volume} {7}},\ \bibinfo {pages} {1020} (\bibinfo {year} {1973})}\BibitemShut {NoStop}%
\bibitem [{\citenamefont {Hirsch}\ and\ \citenamefont {Marsiglio}(1989)}]{Hirsch1}%
  \BibitemOpen
  \bibfield  {author} {\bibinfo {author} {\bibfnamefont {J.~E.}\ \bibnamefont {Hirsch}}\ and\ \bibinfo {author} {\bibfnamefont {F.}~\bibnamefont {Marsiglio}},\ }\href {\doibase 10.1103/PhysRevB.39.11515} {\bibfield  {journal} {\bibinfo  {journal} {Phys. Rev. B}\ }\textbf {\bibinfo {volume} {39}},\ \bibinfo {pages} {11515} (\bibinfo {year} {1989})}\BibitemShut {NoStop}%
\bibitem [{\citenamefont {Marsiglio}\ and\ \citenamefont {Hirsch}(1990)}]{Hirsch2}%
  \BibitemOpen
  \bibfield  {author} {\bibinfo {author} {\bibfnamefont {F.}~\bibnamefont {Marsiglio}}\ and\ \bibinfo {author} {\bibfnamefont {J.~E.}\ \bibnamefont {Hirsch}},\ }\href {\doibase 10.1103/PhysRevB.41.6435} {\bibfield  {journal} {\bibinfo  {journal} {Phys. Rev. B}\ }\textbf {\bibinfo {volume} {41}},\ \bibinfo {pages} {6435} (\bibinfo {year} {1990})}\BibitemShut {NoStop}%
\bibitem [{\citenamefont {Hirsch}\ and\ \citenamefont {Marsiglio}(2019)}]{Hirsch3}%
  \BibitemOpen
  \bibfield  {author} {\bibinfo {author} {\bibfnamefont {J.}~\bibnamefont {Hirsch}}\ and\ \bibinfo {author} {\bibfnamefont {F.}~\bibnamefont {Marsiglio}},\ }\href {\doibase https://doi.org/10.1016/j.physc.2019.1353534} {\bibfield  {journal} {\bibinfo  {journal} {Physica C: Superconductivity and its Applications}\ }\textbf {\bibinfo {volume} {566}},\ \bibinfo {pages} {1353534} (\bibinfo {year} {2019})}\BibitemShut {NoStop}%
\bibitem [{\citenamefont {Anderson}(1987)}]{Anderson}%
  \BibitemOpen
  \bibfield  {author} {\bibinfo {author} {\bibfnamefont {P.~W.}\ \bibnamefont {Anderson}},\ }\href {\doibase 10.1126/science.235.4793.1196} {\bibfield  {journal} {\bibinfo  {journal} {Science}\ }\textbf {\bibinfo {volume} {235}},\ \bibinfo {pages} {1196} (\bibinfo {year} {1987})},\ \Eprint {http://arxiv.org/abs/https://www.science.org/doi/pdf/10.1126/science.235.4793.1196} {https://www.science.org/doi/pdf/10.1126/science.235.4793.1196} \BibitemShut {NoStop}%
\bibitem [{\citenamefont {Bednorz}(1986)}]{cuprate}%
  \BibitemOpen
  \bibfield  {author} {\bibinfo {author} {\bibfnamefont {M.~K.}\ \bibnamefont {Bednorz}, \bibfnamefont {J.G.}},\ }\href {https://doi.org/10.1007/BF01303701} {\bibfield  {journal} {\bibinfo  {journal} {Z. Physik B - Condensed Matter}\ }\textbf {\bibinfo {volume} {64}},\ \bibinfo {pages} {189–193} (\bibinfo {year} {1986})}\BibitemShut {NoStop}%
\bibitem [{\citenamefont {Kung}\ \emph {et~al.}(2016)\citenamefont {Kung}, \citenamefont {Chen}, \citenamefont {Wang}, \citenamefont {Huang}, \citenamefont {Nowadnick}, \citenamefont {Moritz}, \citenamefont {Scalettar}, \citenamefont {Johnston},\ and\ \citenamefont {Devereaux}}]{SJ1}%
  \BibitemOpen
  \bibfield  {author} {\bibinfo {author} {\bibfnamefont {Y.~F.}\ \bibnamefont {Kung}}, \bibinfo {author} {\bibfnamefont {C.-C.}\ \bibnamefont {Chen}}, \bibinfo {author} {\bibfnamefont {Y.}~\bibnamefont {Wang}}, \bibinfo {author} {\bibfnamefont {E.~W.}\ \bibnamefont {Huang}}, \bibinfo {author} {\bibfnamefont {E.~A.}\ \bibnamefont {Nowadnick}}, \bibinfo {author} {\bibfnamefont {B.}~\bibnamefont {Moritz}}, \bibinfo {author} {\bibfnamefont {R.~T.}\ \bibnamefont {Scalettar}}, \bibinfo {author} {\bibfnamefont {S.}~\bibnamefont {Johnston}}, \ and\ \bibinfo {author} {\bibfnamefont {T.~P.}\ \bibnamefont {Devereaux}},\ }\href {\doibase 10.1103/PhysRevB.93.155166} {\bibfield  {journal} {\bibinfo  {journal} {Phys. Rev. B}\ }\textbf {\bibinfo {volume} {93}},\ \bibinfo {pages} {155166} (\bibinfo {year} {2016})}\BibitemShut {NoStop}%
\bibitem [{\citenamefont {Dee}\ \emph {et~al.}(2019)\citenamefont {Dee}, \citenamefont {Nakatsukasa}, \citenamefont {Wang},\ and\ \citenamefont {Johnston}}]{SJ1b}%
  \BibitemOpen
  \bibfield  {author} {\bibinfo {author} {\bibfnamefont {P.~M.}\ \bibnamefont {Dee}}, \bibinfo {author} {\bibfnamefont {K.}~\bibnamefont {Nakatsukasa}}, \bibinfo {author} {\bibfnamefont {Y.}~\bibnamefont {Wang}}, \ and\ \bibinfo {author} {\bibfnamefont {S.}~\bibnamefont {Johnston}},\ }\href {\doibase 10.1103/PhysRevB.99.024514} {\bibfield  {journal} {\bibinfo  {journal} {Phys. Rev. B}\ }\textbf {\bibinfo {volume} {99}},\ \bibinfo {pages} {024514} (\bibinfo {year} {2019})}\BibitemShut {NoStop}%
\bibitem [{\citenamefont {Han}\ \emph {et~al.}(2020)\citenamefont {Han}, \citenamefont {Kivelson},\ and\ \citenamefont {Yao}}]{SJ2}%
  \BibitemOpen
  \bibfield  {author} {\bibinfo {author} {\bibfnamefont {Z.}~\bibnamefont {Han}}, \bibinfo {author} {\bibfnamefont {S.~A.}\ \bibnamefont {Kivelson}}, \ and\ \bibinfo {author} {\bibfnamefont {H.}~\bibnamefont {Yao}},\ }\href {\doibase 10.1103/PhysRevLett.125.167001} {\bibfield  {journal} {\bibinfo  {journal} {Phys. Rev. Lett.}\ }\textbf {\bibinfo {volume} {125}},\ \bibinfo {pages} {167001} (\bibinfo {year} {2020})}\BibitemShut {NoStop}%
\bibitem [{\citenamefont {Wang}\ \emph {et~al.}(2020)\citenamefont {Wang}, \citenamefont {Esterlis}, \citenamefont {Shi}, \citenamefont {Cirac},\ and\ \citenamefont {Demler}}]{SJ3}%
  \BibitemOpen
  \bibfield  {author} {\bibinfo {author} {\bibfnamefont {Y.}~\bibnamefont {Wang}}, \bibinfo {author} {\bibfnamefont {I.}~\bibnamefont {Esterlis}}, \bibinfo {author} {\bibfnamefont {T.}~\bibnamefont {Shi}}, \bibinfo {author} {\bibfnamefont {J.~I.}\ \bibnamefont {Cirac}}, \ and\ \bibinfo {author} {\bibfnamefont {E.}~\bibnamefont {Demler}},\ }\href {\doibase 10.1103/PhysRevResearch.2.043258} {\bibfield  {journal} {\bibinfo  {journal} {Phys. Rev. Res.}\ }\textbf {\bibinfo {volume} {2}},\ \bibinfo {pages} {043258} (\bibinfo {year} {2020})}\BibitemShut {NoStop}%
\bibitem [{\citenamefont {Sawatzky}\ \emph {et~al.}(2009)\citenamefont {Sawatzky}, \citenamefont {Elfimov}, \citenamefont {van~den Brink},\ and\ \citenamefont {Zaanen}}]{george1}%
  \BibitemOpen
  \bibfield  {author} {\bibinfo {author} {\bibfnamefont {G.~A.}\ \bibnamefont {Sawatzky}}, \bibinfo {author} {\bibfnamefont {I.~S.}\ \bibnamefont {Elfimov}}, \bibinfo {author} {\bibfnamefont {J.}~\bibnamefont {van~den Brink}}, \ and\ \bibinfo {author} {\bibfnamefont {J.}~\bibnamefont {Zaanen}},\ }\href {\doibase 10.1209/0295-5075/86/17006} {\bibfield  {journal} {\bibinfo  {journal} {Europhysics Letters}\ }\textbf {\bibinfo {volume} {86}},\ \bibinfo {pages} {17006} (\bibinfo {year} {2009})}\BibitemShut {NoStop}%
\bibitem [{\citenamefont {Berciu}\ \emph {et~al.}(2009)\citenamefont {Berciu}, \citenamefont {Elfimov},\ and\ \citenamefont {Sawatzky}}]{george2}%
  \BibitemOpen
  \bibfield  {author} {\bibinfo {author} {\bibfnamefont {M.}~\bibnamefont {Berciu}}, \bibinfo {author} {\bibfnamefont {I.}~\bibnamefont {Elfimov}}, \ and\ \bibinfo {author} {\bibfnamefont {G.~A.}\ \bibnamefont {Sawatzky}},\ }\href {\doibase 10.1103/PhysRevB.79.214507} {\bibfield  {journal} {\bibinfo  {journal} {Phys. Rev. B}\ }\textbf {\bibinfo {volume} {79}},\ \bibinfo {pages} {214507} (\bibinfo {year} {2009})}\BibitemShut {NoStop}%
\bibitem [{\citenamefont {Valmispild}\ \emph {et~al.}(2020)\citenamefont {Valmispild}, \citenamefont {Dutreix}, \citenamefont {Eckstein}, \citenamefont {Katsnelson}, \citenamefont {Lichtenstein},\ and\ \citenamefont {Stepanov}}]{valmispild}%
  \BibitemOpen
  \bibfield  {author} {\bibinfo {author} {\bibfnamefont {V.~N.}\ \bibnamefont {Valmispild}}, \bibinfo {author} {\bibfnamefont {C.}~\bibnamefont {Dutreix}}, \bibinfo {author} {\bibfnamefont {M.}~\bibnamefont {Eckstein}}, \bibinfo {author} {\bibfnamefont {M.~I.}\ \bibnamefont {Katsnelson}}, \bibinfo {author} {\bibfnamefont {A.~I.}\ \bibnamefont {Lichtenstein}}, \ and\ \bibinfo {author} {\bibfnamefont {E.~A.}\ \bibnamefont {Stepanov}},\ }\href {\doibase 10.1103/PhysRevB.102.220301} {\bibfield  {journal} {\bibinfo  {journal} {Phys. Rev. B}\ }\textbf {\bibinfo {volume} {102}},\ \bibinfo {pages} {220301} (\bibinfo {year} {2020})}\BibitemShut {NoStop}%
\bibitem [{\citenamefont {Tancogne-Dejean}\ \emph {et~al.}(2018)\citenamefont {Tancogne-Dejean}, \citenamefont {Sentef},\ and\ \citenamefont {Rubio}}]{tancogne-dejean}%
  \BibitemOpen
  \bibfield  {author} {\bibinfo {author} {\bibfnamefont {N.}~\bibnamefont {Tancogne-Dejean}}, \bibinfo {author} {\bibfnamefont {M.~A.}\ \bibnamefont {Sentef}}, \ and\ \bibinfo {author} {\bibfnamefont {A.}~\bibnamefont {Rubio}},\ }\href {\doibase 10.1103/PhysRevLett.121.097402} {\bibfield  {journal} {\bibinfo  {journal} {Phys. Rev. Lett.}\ }\textbf {\bibinfo {volume} {121}},\ \bibinfo {pages} {097402} (\bibinfo {year} {2018})}\BibitemShut {NoStop}%
\bibitem [{\citenamefont {Gole\ifmmode~\check{z}\else \v{z}\fi{}}\ \emph {et~al.}(2015)\citenamefont {Gole\ifmmode~\check{z}\else \v{z}\fi{}}, \citenamefont {Eckstein},\ and\ \citenamefont {Werner}}]{golez}%
  \BibitemOpen
  \bibfield  {author} {\bibinfo {author} {\bibfnamefont {D.}~\bibnamefont {Gole\ifmmode~\check{z}\else \v{z}\fi{}}}, \bibinfo {author} {\bibfnamefont {M.}~\bibnamefont {Eckstein}}, \ and\ \bibinfo {author} {\bibfnamefont {P.}~\bibnamefont {Werner}},\ }\href {\doibase 10.1103/PhysRevB.92.195123} {\bibfield  {journal} {\bibinfo  {journal} {Phys. Rev. B}\ }\textbf {\bibinfo {volume} {92}},\ \bibinfo {pages} {195123} (\bibinfo {year} {2015})}\BibitemShut {NoStop}%
\bibitem [{\citenamefont {Baykusheva}\ \emph {et~al.}(2022)\citenamefont {Baykusheva}, \citenamefont {Jang}, \citenamefont {Husain}, \citenamefont {Lee}, \citenamefont {TenHuisen}, \citenamefont {Zhou}, \citenamefont {Park}, \citenamefont {Kim}, \citenamefont {Kim}, \citenamefont {Kim}, \citenamefont {Kim}, \citenamefont {Park}, \citenamefont {Abbamonte}, \citenamefont {Kim}, \citenamefont {Gu}, \citenamefont {Wang},\ and\ \citenamefont {Mitrano}}]{baykusheva}%
  \BibitemOpen
  \bibfield  {author} {\bibinfo {author} {\bibfnamefont {D.~R.}\ \bibnamefont {Baykusheva}}, \bibinfo {author} {\bibfnamefont {H.}~\bibnamefont {Jang}}, \bibinfo {author} {\bibfnamefont {A.~A.}\ \bibnamefont {Husain}}, \bibinfo {author} {\bibfnamefont {S.}~\bibnamefont {Lee}}, \bibinfo {author} {\bibfnamefont {S.~F.~R.}\ \bibnamefont {TenHuisen}}, \bibinfo {author} {\bibfnamefont {P.}~\bibnamefont {Zhou}}, \bibinfo {author} {\bibfnamefont {S.}~\bibnamefont {Park}}, \bibinfo {author} {\bibfnamefont {H.}~\bibnamefont {Kim}}, \bibinfo {author} {\bibfnamefont {J.-K.}\ \bibnamefont {Kim}}, \bibinfo {author} {\bibfnamefont {H.-D.}\ \bibnamefont {Kim}}, \bibinfo {author} {\bibfnamefont {M.}~\bibnamefont {Kim}}, \bibinfo {author} {\bibfnamefont {S.-Y.}\ \bibnamefont {Park}}, \bibinfo {author} {\bibfnamefont {P.}~\bibnamefont {Abbamonte}}, \bibinfo {author} {\bibfnamefont {B.~J.}\ \bibnamefont {Kim}}, \bibinfo {author} {\bibfnamefont {G.~D.}\ \bibnamefont {Gu}}, \bibinfo {author} {\bibfnamefont {Y.}~\bibnamefont
  {Wang}}, \ and\ \bibinfo {author} {\bibfnamefont {M.}~\bibnamefont {Mitrano}},\ }\href {\doibase 10.1103/PhysRevX.12.011013} {\bibfield  {journal} {\bibinfo  {journal} {Phys. Rev. X}\ }\textbf {\bibinfo {volume} {12}},\ \bibinfo {pages} {011013} (\bibinfo {year} {2022})}\BibitemShut {NoStop}%
\bibitem [{\citenamefont {Beaulieu}\ \emph {et~al.}(2021)\citenamefont {Beaulieu}, \citenamefont {Dong}, \citenamefont {Tancogne-Dejean}, \citenamefont {Dendzik}, \citenamefont {Pincelli}, \citenamefont {Maklar}, \citenamefont {Xian}, \citenamefont {Sentef}, \citenamefont {Wolf}, \citenamefont {Rubio}, \citenamefont {Rettig},\ and\ \citenamefont {Ernstorfer}}]{beaulieu}%
  \BibitemOpen
  \bibfield  {author} {\bibinfo {author} {\bibfnamefont {S.}~\bibnamefont {Beaulieu}}, \bibinfo {author} {\bibfnamefont {S.}~\bibnamefont {Dong}}, \bibinfo {author} {\bibfnamefont {N.}~\bibnamefont {Tancogne-Dejean}}, \bibinfo {author} {\bibfnamefont {M.}~\bibnamefont {Dendzik}}, \bibinfo {author} {\bibfnamefont {T.}~\bibnamefont {Pincelli}}, \bibinfo {author} {\bibfnamefont {J.}~\bibnamefont {Maklar}}, \bibinfo {author} {\bibfnamefont {R.~P.}\ \bibnamefont {Xian}}, \bibinfo {author} {\bibfnamefont {M.~A.}\ \bibnamefont {Sentef}}, \bibinfo {author} {\bibfnamefont {M.}~\bibnamefont {Wolf}}, \bibinfo {author} {\bibfnamefont {A.}~\bibnamefont {Rubio}}, \bibinfo {author} {\bibfnamefont {L.}~\bibnamefont {Rettig}}, \ and\ \bibinfo {author} {\bibfnamefont {R.}~\bibnamefont {Ernstorfer}},\ }\href {\doibase 10.1126/sciadv.abd9275} {\bibfield  {journal} {\bibinfo  {journal} {Science Advances}\ }\textbf {\bibinfo {volume} {7}},\ \bibinfo {pages} {eabd9275} (\bibinfo {year} {2021})}\BibitemShut {NoStop}%
\bibitem [{\citenamefont {Caprara}\ \emph {et~al.}(2017)\citenamefont {Caprara}, \citenamefont {Di~Castro}, \citenamefont {Seibold},\ and\ \citenamefont {Grilli}}]{Caprara}%
  \BibitemOpen
  \bibfield  {author} {\bibinfo {author} {\bibfnamefont {S.}~\bibnamefont {Caprara}}, \bibinfo {author} {\bibfnamefont {C.}~\bibnamefont {Di~Castro}}, \bibinfo {author} {\bibfnamefont {G.}~\bibnamefont {Seibold}}, \ and\ \bibinfo {author} {\bibfnamefont {M.}~\bibnamefont {Grilli}},\ }\href {\doibase 10.1103/PhysRevB.95.224511} {\bibfield  {journal} {\bibinfo  {journal} {Phys. Rev. B}\ }\textbf {\bibinfo {volume} {95}},\ \bibinfo {pages} {224511} (\bibinfo {year} {2017})}\BibitemShut {NoStop}%
\bibitem [{\citenamefont {Boschini}\ \emph {et~al.}(2021)\citenamefont {Boschini}, \citenamefont {Minola}, \citenamefont {Sutarto}, \citenamefont {Schierle}, \citenamefont {Bluschke}, \citenamefont {Das}, \citenamefont {Yang}, \citenamefont {Michiardi}, \citenamefont {Shao}, \citenamefont {Feng}, \citenamefont {Ono}, \citenamefont {Zhong}, \citenamefont {Schneeloch}, \citenamefont {Gu}, \citenamefont {Weschke}, \citenamefont {He}, \citenamefont {Chuang}, \citenamefont {Keimer}, \citenamefont {Damascelli}, \citenamefont {Frano},\ and\ \citenamefont {da~Silva~Neto}}]{Boschini}%
  \BibitemOpen
  \bibfield  {author} {\bibinfo {author} {\bibfnamefont {F.}~\bibnamefont {Boschini}}, \bibinfo {author} {\bibfnamefont {M.}~\bibnamefont {Minola}}, \bibinfo {author} {\bibfnamefont {R.}~\bibnamefont {Sutarto}}, \bibinfo {author} {\bibfnamefont {E.}~\bibnamefont {Schierle}}, \bibinfo {author} {\bibfnamefont {M.}~\bibnamefont {Bluschke}}, \bibinfo {author} {\bibfnamefont {S.}~\bibnamefont {Das}}, \bibinfo {author} {\bibfnamefont {Y.}~\bibnamefont {Yang}}, \bibinfo {author} {\bibfnamefont {M.}~\bibnamefont {Michiardi}}, \bibinfo {author} {\bibfnamefont {Y.~C.}\ \bibnamefont {Shao}}, \bibinfo {author} {\bibfnamefont {X.}~\bibnamefont {Feng}}, \bibinfo {author} {\bibfnamefont {S.}~\bibnamefont {Ono}}, \bibinfo {author} {\bibfnamefont {R.~D.}\ \bibnamefont {Zhong}}, \bibinfo {author} {\bibfnamefont {J.~A.}\ \bibnamefont {Schneeloch}}, \bibinfo {author} {\bibfnamefont {G.~D.}\ \bibnamefont {Gu}}, \bibinfo {author} {\bibfnamefont {E.}~\bibnamefont {Weschke}}, \bibinfo {author} {\bibfnamefont {F.}~\bibnamefont {He}},
  \bibinfo {author} {\bibfnamefont {Y.~D.}\ \bibnamefont {Chuang}}, \bibinfo {author} {\bibfnamefont {B.}~\bibnamefont {Keimer}}, \bibinfo {author} {\bibfnamefont {A.}~\bibnamefont {Damascelli}}, \bibinfo {author} {\bibfnamefont {A.}~\bibnamefont {Frano}}, \ and\ \bibinfo {author} {\bibfnamefont {E.~H.}\ \bibnamefont {da~Silva~Neto}},\ }\href {\doibase 10.1038/s41467-020-20824-7} {\bibfield  {journal} {\bibinfo  {journal} {Nature Communications}\ }\textbf {\bibinfo {volume} {12}} (\bibinfo {year} {2021}),\ 10.1038/s41467-020-20824-7}\BibitemShut {NoStop}%
\bibitem [{\citenamefont {Giannetti}\ \emph {et~al.}(2016)\citenamefont {Giannetti}, \citenamefont {Capone}, \citenamefont {Fausti}, \citenamefont {Fabrizio}, \citenamefont {Parmigiani},\ and\ \citenamefont {Mihailovic}}]{N1}%
  \BibitemOpen
  \bibfield  {author} {\bibinfo {author} {\bibfnamefont {C.}~\bibnamefont {Giannetti}}, \bibinfo {author} {\bibfnamefont {M.}~\bibnamefont {Capone}}, \bibinfo {author} {\bibfnamefont {D.}~\bibnamefont {Fausti}}, \bibinfo {author} {\bibfnamefont {M.}~\bibnamefont {Fabrizio}}, \bibinfo {author} {\bibfnamefont {F.}~\bibnamefont {Parmigiani}}, \ and\ \bibinfo {author} {\bibfnamefont {D.}~\bibnamefont {Mihailovic}},\ }\href {\doibase 10.1080/00018732.2016.1194044} {\bibfield  {journal} {\bibinfo  {journal} {Advances in Physics}\ }\textbf {\bibinfo {volume} {65}},\ \bibinfo {pages} {58} (\bibinfo {year} {2016})}\BibitemShut {NoStop}%
\bibitem [{\citenamefont {Mitrano}\ and\ \citenamefont {Wang}(2020)}]{mitrano2020probing}%
  \BibitemOpen
  \bibfield  {author} {\bibinfo {author} {\bibfnamefont {M.}~\bibnamefont {Mitrano}}\ and\ \bibinfo {author} {\bibfnamefont {Y.}~\bibnamefont {Wang}},\ }\href@noop {} {\bibfield  {journal} {\bibinfo  {journal} {Communications Physics}\ }\textbf {\bibinfo {volume} {3}},\ \bibinfo {pages} {184} (\bibinfo {year} {2020})}\BibitemShut {NoStop}%
\bibitem [{\citenamefont {Cvetkovic}\ and\ \citenamefont {Tesanovic}(2009)}]{MM1}%
  \BibitemOpen
  \bibfield  {author} {\bibinfo {author} {\bibfnamefont {V.}~\bibnamefont {Cvetkovic}}\ and\ \bibinfo {author} {\bibfnamefont {Z.}~\bibnamefont {Tesanovic}},\ }\href {\doibase 10.1209/0295-5075/85/37002} {\bibfield  {journal} {\bibinfo  {journal} {Europhysics Letters}\ }\textbf {\bibinfo {volume} {85}},\ \bibinfo {pages} {37002} (\bibinfo {year} {2009})}\BibitemShut {NoStop}%
\bibitem [{\citenamefont {Andersen}\ and\ \citenamefont {Boeri}(2011)}]{MM2}%
  \BibitemOpen
  \bibfield  {author} {\bibinfo {author} {\bibfnamefont {O.~K.}\ \bibnamefont {Andersen}}\ and\ \bibinfo {author} {\bibfnamefont {L.}~\bibnamefont {Boeri}},\ }\href {\doibase 10.1002/andp.201000149} {\bibfield  {journal} {\bibinfo  {journal} {Annalen der Physik}\ }\textbf {\bibinfo {volume} {523}},\ \bibinfo {pages} {8} (\bibinfo {year} {2011})}\BibitemShut {NoStop}%
\bibitem [{\citenamefont {Haule}\ \emph {et~al.}(2008)\citenamefont {Haule}, \citenamefont {Shim},\ and\ \citenamefont {Kotliar}}]{MM3}%
  \BibitemOpen
  \bibfield  {author} {\bibinfo {author} {\bibfnamefont {K.}~\bibnamefont {Haule}}, \bibinfo {author} {\bibfnamefont {J.~H.}\ \bibnamefont {Shim}}, \ and\ \bibinfo {author} {\bibfnamefont {G.}~\bibnamefont {Kotliar}},\ }\href {\doibase 10.1103/PhysRevLett.100.226402} {\bibfield  {journal} {\bibinfo  {journal} {Phys. Rev. Lett.}\ }\textbf {\bibinfo {volume} {100}},\ \bibinfo {pages} {226402} (\bibinfo {year} {2008})}\BibitemShut {NoStop}%
\bibitem [{\citenamefont {Boschini}\ \emph {et~al.}(2024)\citenamefont {Boschini}, \citenamefont {Zonno},\ and\ \citenamefont {Damascelli}}]{ARPES_review}%
  \BibitemOpen
  \bibfield  {author} {\bibinfo {author} {\bibfnamefont {F.}~\bibnamefont {Boschini}}, \bibinfo {author} {\bibfnamefont {M.}~\bibnamefont {Zonno}}, \ and\ \bibinfo {author} {\bibfnamefont {A.}~\bibnamefont {Damascelli}},\ }\href {\doibase 10.1103/RevModPhys.96.015003} {\bibfield  {journal} {\bibinfo  {journal} {Rev. Mod. Phys.}\ }\textbf {\bibinfo {volume} {96}},\ \bibinfo {pages} {015003} (\bibinfo {year} {2024})}\BibitemShut {NoStop}%
\bibitem [{\citenamefont {Kova\ifmmode~\check{c}\else \v{c}\fi{}}\ \emph {et~al.}(2024)\citenamefont {Kova\ifmmode~\check{c}\else \v{c}\fi{}}, \citenamefont {Gole\ifmmode~\check{z}\else \v{z}\fi{}}, \citenamefont {Mierzejewski},\ and\ \citenamefont {Bon\ifmmode~\check{c}\else \v{c}\fi{}a}}]{Janez}%
  \BibitemOpen
  \bibfield  {author} {\bibinfo {author} {\bibfnamefont {K.}~\bibnamefont {Kova\ifmmode~\check{c}\else \v{c}\fi{}}}, \bibinfo {author} {\bibfnamefont {D.}~\bibnamefont {Gole\ifmmode~\check{z}\else \v{z}\fi{}}}, \bibinfo {author} {\bibfnamefont {M.}~\bibnamefont {Mierzejewski}}, \ and\ \bibinfo {author} {\bibfnamefont {J.}~\bibnamefont {Bon\ifmmode~\check{c}\else \v{c}\fi{}a}},\ }\href {\doibase 10.1103/PhysRevLett.132.106001} {\bibfield  {journal} {\bibinfo  {journal} {Phys. Rev. Lett.}\ }\textbf {\bibinfo {volume} {132}},\ \bibinfo {pages} {106001} (\bibinfo {year} {2024})}\BibitemShut {NoStop}%
\end{thebibliography}%

\end{document}